\RequirePackage{lineno} 

\documentclass[12pt,preprint]{aastex}

\usepackage{natbib}
\usepackage{color}
\usepackage{multirow}

\shorttitle{3D GCM Intercomparison}
\shortauthors{Yang et al.}

\begin{document}

\modulolinenumbers[1]




\title{Simulations of Water Vapor and Clouds on Rapidly Rotating and 
Tidally Locked Planets: A 3D Model Intercomparison}

\author{Jun Yang$^{1,2}$, J\'{e}r\'{e}my Leconte$^3$, Eric
  T. Wolf$^4$, Timothy Merlis$^5$, Daniel D.B. Koll$^6$, 
  Fran\c{c}ois Forget$^7$, and Dorian S. Abbot$^8$} 
  \affil{$^1$Dept. of Atmospheric and Oceanic Sciences, School of Physics, Peking University, Beijing, 100871, China\\  
  $^2$Previously at Dept. of the Geophysical Sciences, University of Chicago, Chicago, IL, 60637, USA\\
  $^3$Laboratoire d'astrophysique de Bordeaux, Univ. Bordeaux, CNRS, B18N, all\'ee Groffroy Saint-Hilaire, Pessac, 33615, France\\ 
  $^4$Lab. for Atmospheric and Space Physics, University of Colorado in Boulder, Colorado, 80303, USA \\ 
  $^5$Dept. of Atmospheric and Oceanic Sciences at McGill University, Montr\'{e}al, QC, H3A 0G4, Canada\\ 
  $^6$Dept. of Earth, Atmospheric and Planetary Sciences, MIT, Cambridge, MA, 02139, USA\\
  $^7$Lab. de M\'et\'eorologie Dynamique, Institut Pierre Simon Laplace, CNRS, Paris, France\\ 
  $^8$Dept. of the Geophysical Sciences, University of Chicago, Chicago, IL, 60637, USA}
\email{Correspondence:~junyang@pku.edu.cn}

\begin{abstract}
Robustly modeling the inner edge of the habitable zone is essential
for determining the most promising potentially habitable exoplanets
for atmospheric characterization. Global Climate Models (GCMs) have
become the standard tool for calculating this boundary, but divergent
results have emerged among the various GCMs. In this study we perform
an intercomparison of standard GCMs used in the field on a rapidly
rotating planet receiving a G-star spectral energy distribution and on
a tidally locked planet receiving an M-star spectral energy
distribution. Experiments both with and without clouds are examined.
We find relatively small difference (within 8~K) in global-mean
surface temperature simulation among the models in the G-star case
with clouds. In contrast, the global-mean surface temperature
simulation in the M-star case is highly divergent (20-30~K). Moreover,
even differences in the simulated surface temperature when clouds are
turned off are significant. These differences are caused
by differences in cloud simulation and/or radiative transfer, as well
as complex interactions between atmospheric dynamics and these two
processes. For example we find that an increase in atmospheric
absorption of shortwave radiation can lead to higher relative humidity
at high altitudes globally and therefore a significant decrease in
planetary radiation emitted to space. This study emphasizes the
importance of basing conclusions about planetary climate on
simulations from a variety of GCMs, and motivates the eventual
comparison of GCM results with terrestrial exoplanet observations to
improve their performance.
\end{abstract}

\keywords{astrobiology --- planets and satellites: atmospheres 
---  planets and satellites: general --- methods: numerical --- radiative transfer}

\section{Introduction}
\label{sec:introduction}

The ``habitable zone'' is the circumstellar region where an Earth-like
planet can support liquid water on its surface
\citep{Kasting93,Kopparapu:2013,kasting2014remote}, which is essential
for Earth-like life. The habitable zone concept has received
increasing attention in recent years as the number of potentially
habitable extrasolar planets has increased and future NASA missions 
to characterize the atmospheres of potentially habitable extrasolar
planets are being planned. This has led to the application of
sophisticated three-dimensional (3D) GCMs, which are capable of modeling
atmospheric dynamics, clouds, and water vapor distributions, to the
problem
\citep[e.g.,][]{Merlis:2010,Edson:2011p2367,Pierrehumbert:2011p3287,
  leconte2013increased,Leconte:2013gv,yang2013,yang2014,shields2013effect,
  shields2014spectrum,Wolf:2014,Wangetal2014:eccentric,
  Wangetal2016:obliquity,wolf2015evolution,
  way2015exploring,way2017resolving,godolt20153d,kopparapu2016inner,popp2016, Carone_paper1,Carone_paper2,Carone_paper3,Carone_paper4,turbet2016habitability,
  salameh2017role,wolf2017assessing,
  wolf2017constraints,haqq2017demarcating,boutle2017exploring,
  kopparapu2017habitable,Lewisetal2018,Binetal2018,turbet2018modeling}.

The inner edge of the habitable zone is marked by either a massive
increase in surface temperature as a result of a fundamental limit on
infrared emission to space by an Earth-like planet, the ``runaway
greenhouse,'' or the loss of a planet's water through
photodissociation and hydrodynamic escape due to high surface
temperatures and a moist stratosphere, a ``moist greenhouse''
\citep{Kasting88}. Since modern Earth is relatively near the inner
edge of the habitable zone \citep{Kopparapu:2013}, aspects of these
processes can be modeled using GCMs that were primarily designed to
model the climate of modern and ancient Earth. In contrast, modeling
the outer edge of the habitable zone requires accurate modeling of
radiative transfer at high CO$_2$ concentrations, CO$_2$ clouds, and
the dynamical effects of CO$_2$ condensation
\citep{Wordsworth:2011p3221,Wordsworth2015,
  turbet2016habitability}. As a result, more GCMs have been applied to
the inner edge of the habitable zone, which has exposed the dramatic
impact of differences in model formulations on its position.

GCMs generally disagree on the position of the inner edge of the habitable zone both for
planets orbiting cool M stars and Sun-like G stars. For example, for a
tidally locked planet orbiting an M star, \citet{kopparapu2017habitable} 
found that for a stellar temperature of 3,400 K, updating the radiative 
scheme in the CAM4 GCM moved the inner edge of the habitable zone
(runaway greenhouse) from 83\,\% above modern Earth's solar constant to
39\,\% above it. Moreover, if we consider a planet with Earth's rotation
rate receiving the Sun's spectral energy distribution, a runaway
greenhouse occurs in the LMDG GCM when the solar constant is increased
by 10\,\% above modern Earth's value \citep{Leconte:2013gv}, but has not
occurred when it is increased by 15\,\% in ECHAM6 \citep{popp2016} and
by 21\,\% in CAM4\_WOLF \citep{wolf2015evolution}. These authors
diagnosed differences in cloud simulation among their GCMs, but the
lack of a uniform modeling framework made it impossible to firmly
establish the cause of differences in cloud behavior, as well as
whether variation in other processes might be important.

To clarify the situation, we organized a GCM intercomparison to
investigate the causes of differences among GCMs that have been used
to simulate the inner edge of the habitable zone in more detail. The
participating GCMs are listed in Table~1. We started with a set of
standardized one-dimensional radiative calculations with assumed
vertical profiles of temperature and water vapor, and found that
differences among the GCM radiative schemes in both longwave and
shortwave are mainly due to differences in water vapor absorption
\citep{Yang2016}. LMDG had the strongest greenhouse (longwave) effect
and CAM3 had the weakest, with 17 W~m$^{-2}$ difference between them
at a surface temperature of 320~K. In shortwave, CAM4\_Wolf was the
most absorptive\footnote{E. T. Wolf has corrected this bias by 
improving the wavelength resolution of the stellar spectrum and 
absorption coefficients \citep[e.g.,][]{wolf2017assessing,wolf2017constraints}.} 
and CAM3 was the least, with a maximum top-of-atmosphere difference of
$\sim$$10$~W~m$^{-2}$ for a G star spectral energy distribution and
$\sim$$20$~W~m$^{-2}$ for a M star spectral energy distribution. The more
sophisticated line-by-line radiative codes fell between these extremes
in both longwave and shortwave. When we combined both longwave and
shortwave fluxes to estimate the effective stellar flux of the inner
edge of the habitable zone, we found a variation of about 10\,\% of
modern Earth's stellar flux among the GCMs due to differences in the
treatment of water vapor radiative transfer alone.

The purpose of this paper is to extend the analysis of
\citet{Yang2016} to the three-dimensional effects of GCMs. We have
performed a set of standardized simulations for continent-free planets
with (1) Earth's rotation rate and the Sun's spectral energy
distribution, and (2) a tidally locked orbital configuration and an
idealized M star spectral energy distribution
(section~\ref{sec:methods}).  We will investigate model variation in
these simulations, as well as analyze additional simulations designed
to identify in more detail differences between CAM3, a cooler GCM, and
LMDG, a warmer GCM (section~\ref{sec:results}). We conclude and discuss
implications of this work in section~\ref{sec:conclusiondiscussion}.


\section{Methods}
\label{sec:methods}

The GCMs studied in this intercomparison and their resolutions are
displayed in Table~1. CAM3 is CAM version 3, developed at NCAR
\citep{Collins-Hack-Boville-et-al-2002:description}.  CAM4 is CAM
version 4 \citep{neale2010description} and CAM4\_Wolf is CAM4 with a
new radiative transfer module (\url{https://wiki.ucar.edu/display/etcam/Extraterrestrial+CAM}, 
see \cite{wolf2015evolution}). AM2 is a 3D GCM developed at 
NOAA/GFDL \citep{gfdl2004new}. LMDG is the 3D Laboratoire de M\'et\'eorologie 
Dynamique (LMD) Generic Model, developed at LMD
\citep{wordsworth2010infrared,wordsworth2010gliese,Wordsworth:2011p3221,Forget2012}.

\begin{table*}[!htbp]
\label{tab:GCMs}
\begin{center}
  \caption{List of the GCMs in this intercomparison. The horizontal
    resolution is given as latitude by longitude. We also performed
    some CAM3 simulations with $2.8^\circ\times 2.8^\circ$ horizontal
    resolution as well as CAM4 simulations with a $1.9^\circ \times
    1.9^\circ$ horizontal resolution and with a finite volume
    dynamical core. Sensitivity tests using CAM3 with different model top 
    pressures (0.9 and 0.1 hPa) showed that the model top 
  pressure does not significantly influence the surface temperature.}
\begin{tabular}{lllll}
  \tableline\tableline\noalign{\smallskip}
  GCM & Resolution & Levels & Top Pressure & Dynamical Core \\
  \tableline\noalign{\smallskip}
CAM3  & $3.75^\circ \times 3.75^\circ$ & 26 & 3.0 hPa & Spectral \\
CAM4  & $3.75^\circ \times 3.75^\circ$ & 26 &3.0 hPa & Spectral \\
CAM4\_Wolf & $4.0^\circ \times 5.0^\circ$ & 45 & 0.2 hPa & Spectral\\
AM2   & $2.0^\circ \times 2.0^\circ$ & 32 & 2.2 hPa & Finite-volume  \\
LMDG  & $2.8^\circ \times 2.8^\circ$ & 30 & 1.0 hPa & Finite-difference \\
  \tableline\noalign{\smallskip}
  \end{tabular}
  \end{center}
\end{table*}

We ran each GCM in a standard set of conditions. First, we ran them
with a G star spectral energy distribution, a 24 hour rotation period,
365 Earth days per year, and both with and without clouds. Second, we
ran them with an M star spectral energy distribution, tidally locked
in a 1:1 synchronous rotation state, with a 60 day orbital and
rotation period, and both with and without clouds. The G star spectral
energy distribution was the default distribution in the GCM for the
Sun. The M star spectral energy distribution was a blackbody Planck
distribution corresponding to a temperature of 3,400~K.  AM2's M star,
with clouds experiment did not converge due to an unresolved problem,
so it is not listed in the following figures or tables.

We used a stellar flux of 1,360~W~m$^{-2}$, zero obliquity, zero
eccentricity and Earth's radius and gravity for all simulations. We
ran the models in aqua-planet mode (with no continents) with a 50-m
deep mixed layer ocean and no ocean heat transport. The atmosphere was
1 bar of N$_2$ with 376~ppmv CO$_2$ and a variable amount of
H$_2$O. We set CH$_4$, N$_2$O, CFCs, O$_3$, O$_2$, and all aerosols to
zero. We assumed no snow or sea ice, but allowed the sea surface
temperature to drop below the freezing point.  We set the surface
albedo to 0.05 everywhere. We performed simulations both with clouds
set to zero (more exactly, the radiative effects of clouds were turned
off but cloud water and precipitation still formed) and with clouds
calculated by the GCM cloud schemes.  It should be noted here that
because of the diversity of environments that have been modeled with
LMDG, there are several possible cloud parametrizations
available. Here we use the cloud parametrization and
parameters from \cite{charnay2013exploring}, where the cloud particle
size distributions for both liquid droplets and ice particles are
fixed. Another important point is how the total cloud fraction of an
atmospheric column --- the one that will be used in the radiative
transfer calculation --- is computed from the cloud fractions at all
the modeled altitudes.  In our baseline run (LMDG\_max), we assume
that clouds have a maximal recovery probability so that the total
cloud fraction of the column is equal to the maximum cloud fraction at
any altitude. We also present another set of simulations where we make
the assumption that clouds at each level are uncorrelated, resulting
in a random overlap (LMDG\_random). We also performed more detailed
simulations using CAM3 and LMDG run at a variety of stellar fluxes,
and one example where we set water vapor to zero (a dry atmosphere) in
both of these two models.


\section{Results}
\label{sec:results}

\subsection{G-star Planet GCM Comparison}
\label{sec:G-Star-Spectrum}

Our first comparison involved planets in an orbital and rotational
configuration similar to modern Earth's, exposed to a G-star spectrum,
with cloud radiative effects set to zero. As might be expected,
differences in planetary albedo are small, within 0.01 (Table 2). This
is consistent with the fact that the planetary albedos were fairly
similar when the GCMs were run in 1D radiative-convective mode and
forced with a G-star spectrum \citep[see Fig.~7 in][]{Yang2016}. Part
of the explanation for this may also be that the surface albedo is very low
(0.05) in our experiments, so that 95\% of light hitting the surface
is absorbed, and differences among the models in shortwave absorption
by atmospheric water vapor can have less effect on the planetary
albedo. \cite{Yang2016} have already pointed out significant
differences in longwave radiative transfer that could result in
different surface temperatures for a given shortwave heating, and this
trend is confirmed here. Global-mean surface temperatures are within 1
K among CAM3, AM2, and CAM4\_Wolf, but are 6 K lower in CAM4 and
$\approx$12 K higher in LMDG, which was the model with the most pronounced
greenhouse effect of water vapor \citep[Fig.~3(a) in][]{Yang2016}.

It should not be too surprising to find such significant differences
in temperature simulation of the group without clouds, especially between LMDG and
the other GCMs. Turning off clouds in our aqua-planet configuration
with a surface albedo of 0.05 results in a bond albedo of
$\approx$0.10-0.11, or equivalently a mean absorbed stellar flux of
$\approx$303-306 W\,m$^{-2}$. This is just at the limit where some 1D
saturated radiative-convective models are in a runaway greenhouse,
such as LMDG and CAM4\_Wolf, and others are not, such as CAM3 and CAM4
(see Fig.~3(a) of \cite{Yang2016}).  Fortunately, atmospheric
circulation--induced sub-saturation in the substropics makes all the
models in this experiment stable
\citep{Pierrehumbert-1995:thermostats,leconte2013increased}, but they
are still functioning in a regime of high climate sensitivity due to
the strong positive water vapor feedback. This means that
small variations in shortwave absorption can lead to large variations
in surface temperature.

\begin{figure}[!htbp]
\begin{center}
\includegraphics[angle=0, width=14cm]{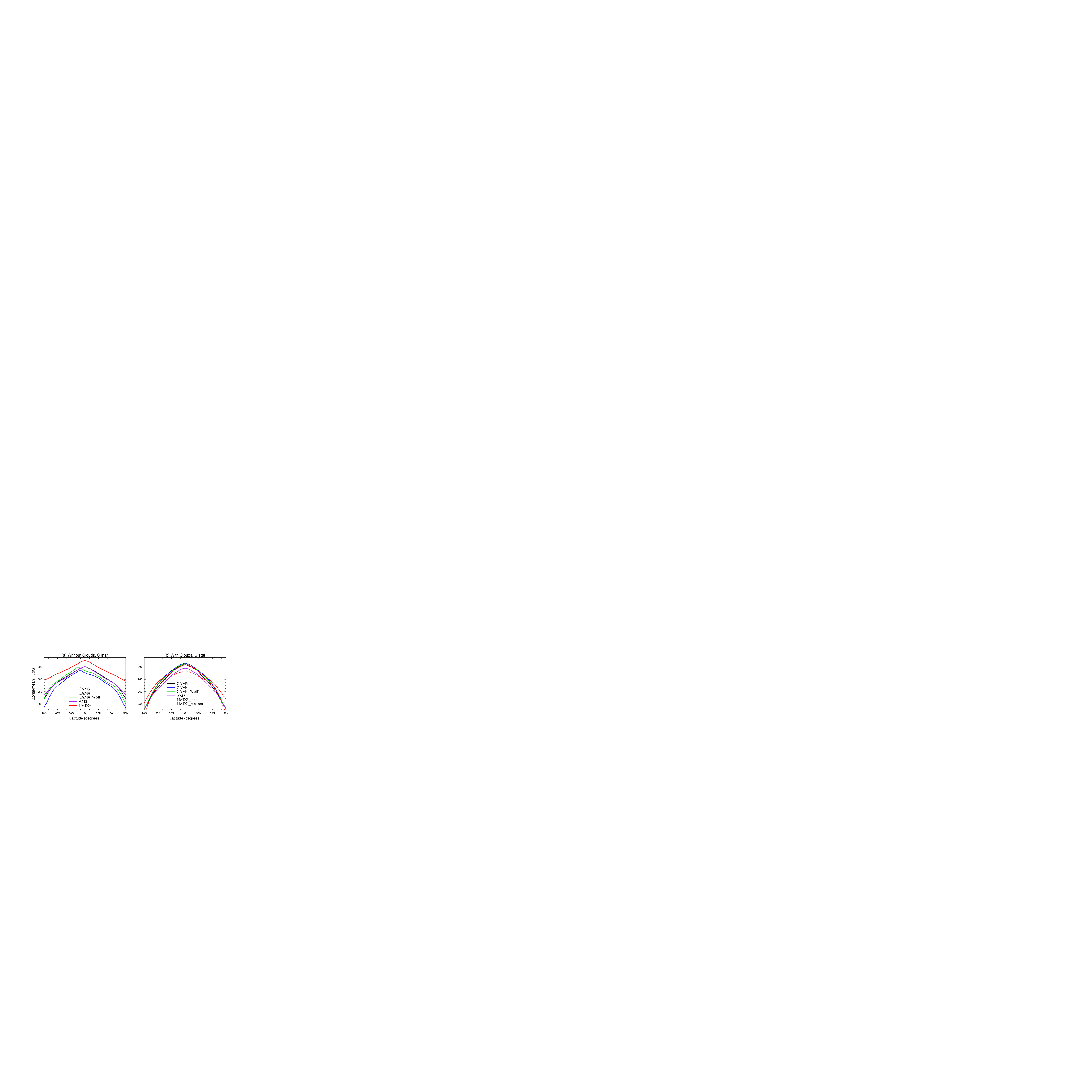}
\caption{G star surface temperature: Zonal (East-West) mean surface
  temperature as a function of latitude for all GCMs both without
  clouds (a) and including clouds (b). The simulations assume
  a rapidly rotating aqua-planet with a G star stellar spectrum and a
  stellar flux of 1,360~W~m$^{-2}$. LMDG\_max assumes maximum overlap 
  between different types of clouds at each altitude while LMDG\_random 
  employs a random overlap. Note the different y axis ranges between (a) and (b).}
\label{fig:GStarComparison}
\end{center}
\end{figure}

Interestingly, both CAM4 and CAM4\_Wolf exhibit spontaneous symmetry
breaking in the cloud-free configuration, with a meridionally
asymmetric climate resulting from symmetric boundary conditions
(Fig.~\ref{fig:GStarComparison}(a)). Sensitivity tests using CAM4\_Wolf
show that the hemisphere that contains the maximum in surface
temperature depends on the initial conditions (figure not shown). The asymmetry results
in a climate that is cooler than the other GCMs (Table~2),
particularly in the case of CAM4, and foreshadows the important
effects that differences in the simulation of atmospheric dynamics can
produce in model climates in certain situations, which we will
investigate further in section~\ref{sec:explaining}.

\begin{table}[!htbp]  
\label{tab:GstarTable}
\begin{center}
\caption{Global-mean climatic characteristics of the G-star spectrum GCM simulations.}
\begin{tabular}{llcccccccc}
  \tableline\tableline\noalign{\smallskip}
Simulations  & GCMs & T$_S$$^{a}$  &  $\alpha_p$$^{b}$  & SWCE$^{c}$ & LWCE$^{d}$ & NCE$^{e}$ & Cld$^{f}$  & WVP$^{g}$ & CWP$^{h}$ \\
          & &  [K]  &   [0-1]   & [Wm$^{-2}$] & [Wm$^{-2}$] & [Wm$^{-2}$] & [\%]  & [kgm$^{-2}$]  & [kgm$^{-2}$] \\
 \tableline\noalign{\smallskip}
\multirow{5}{*}{No Clouds}  & CAM3   & 307       & 0.10    & --- & ---&  ---& --- & 100  & ---     \\
\multirow{5}{*}{}            & CAM4   & 301       & 0.10   & ---& ---&---  &  --- & 62   & ---  \\
\multirow{5}{*}{}              & CAM4\_Wolf   &  306     & 0.11 &--- & ---&  ---&  ---  & 95   & ---    \\
\multirow{5}{*}{}             & AM2     & 307      & 0.10   & ---& ---&  ---& --- & 99   & ---    \\
\multirow{5}{*}{}             & LMDG   & 319     &  0.11  & ---&--- &  ---& --- & 178   & ---  \\
 \tableline\noalign{\smallskip}
\multirow{6}{*}{With Clouds}   & CAM3 & 287    & 0.33    & --78  & 39  & --39 & 77 & 25 & 0.18 \\
\multirow{6}{*}{}   & CAM4 & 290    & 0.32    & --73  & 41  & --32 & 70 & 34               &  0.20 \\
\multirow{6}{*}{}   & CAM4\_Wolf    & 289     & 0.33  & --74  & 34   & --40 & 70 & 30  & 0.17 \\
\multirow{6}{*}{}   & AM2                 & 282     & 0.35  & --83  & 35  & --48  & 86  & 15 &  0.08 \\
\multirow{6}{*}{}   & LMDG\_max       & 290     & 0.30   & --64  & 30  & --34 & 43 & 24 & 0.15\\
\multirow{6}{*}{}  & LMDG\_random   &  282   & 0.39    & --94 & 39  & --55  & 86 & 12 & 0.13 \\
 \tableline\noalign{\smallskip}
\end{tabular}
\emph{a.}~T$_S$: global-mean surface temperature \\
\emph{b.}~$\alpha_p$: planetary albedo\\
\emph{c.}~SWCE: shortwave cloud radiative effect at the top of the model\\
\emph{d.}~LWCE: longwave cloud radiative effect at the top of the model\\
\emph{e.}~NCE: SWCE + LWCE \\
\emph{f.}~Cld: the total cloud coverage \\
\emph{g.}~WVP: the vertical-integrated water vapor content\\
\emph{h.}~CWP: the vertical-integrated cloud water (liquid plus ice) content
\end{center}

\end{table}

\begin{figure}[!htbp]
\begin{center}
\includegraphics[angle=0, width=16.5cm]{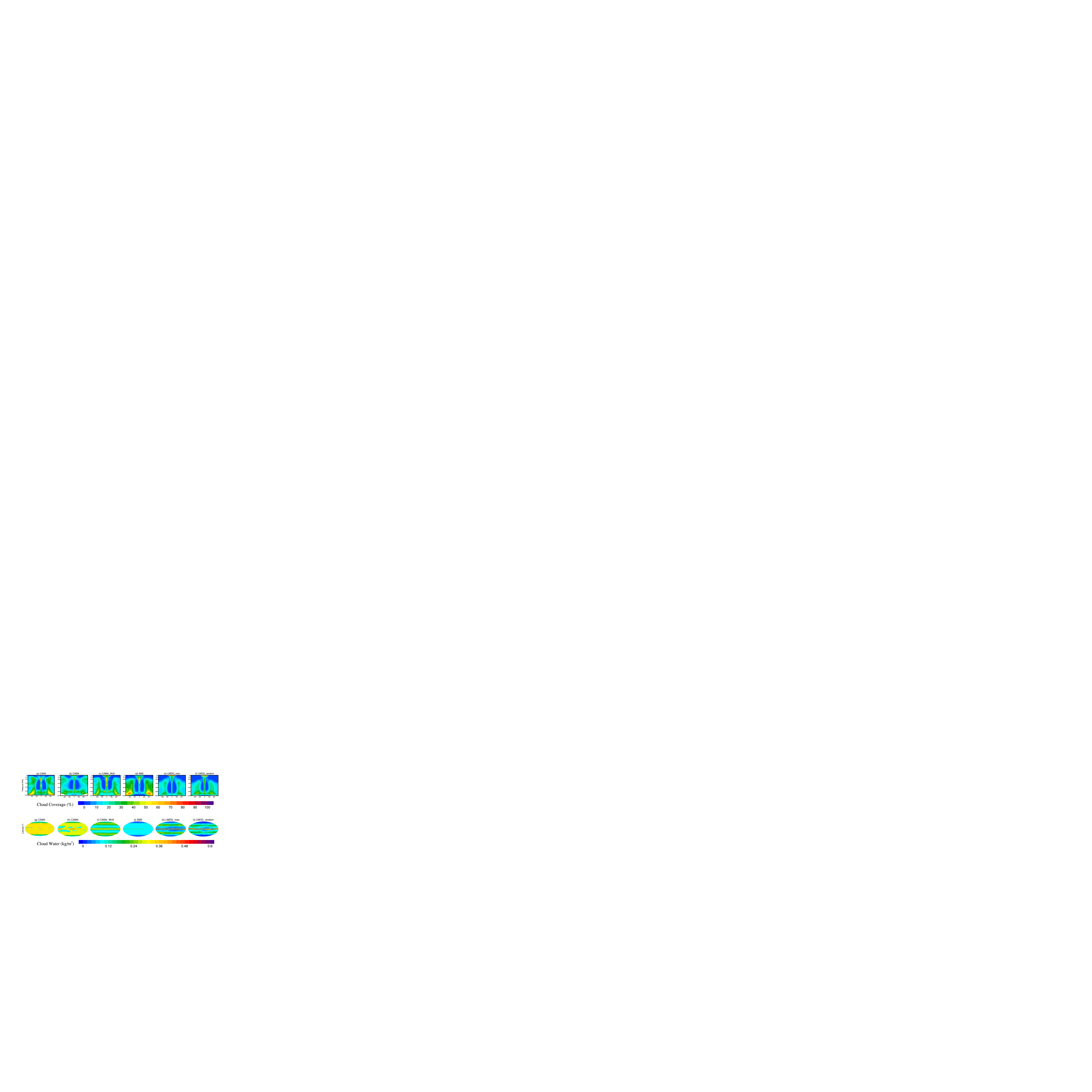}
\caption{G star clouds: Upper panels: Contour plots of zonal (East-West) mean cloud
  fraction as a function of latitude and pressure (vertical) and lower panels: Vertically integrated 
  cloud water amount (including both liquid and ice phases), for all
  GCMs. The simulations assume a rapidly rotating aqua-planet with a G
  star stellar spectrum and a stellar flux of 1,360~W~m$^{-2}$.}
\label{fig:GStarClouds}
\end{center}
\end{figure}

Including clouds cools all models (Table~2 and
Fig.~\ref{fig:GStarComparison}(b)), which is expected since clouds
cool modern Earth. The global-mean net cloud radiative effect among
the models varies greatly, from $-$32 to $-$55~W~m$^{-2}$, mainly due
to differences in cloud fraction and cloud water amount
parameterizations (Fig.~\ref{fig:GStarClouds}). The cloud radiative
effect is more negative than its value on modern Earth
\citep[$-$20~W~m$^{-2}$,][]{kiehl1997earth}, mainly because of the low
surface albedo of a continent-free planet. Also, when clouds are
included in the G-star spectrum calculations, meridional symmetry is
restored to both CAM4 and CAM4\_Wolf
(Fig.~\ref{fig:GStarComparison}(b)).

Interestingly, the global-mean surface temperature is more similar
among the models when clouds are included than when they are not
(Table~2). Part of the explanation for this may be that the models are
cooler when clouds are included, and therefore farther from the
runaway greenhouse where the climate sensitivity is
high. Additionally, all of the models have been tuned to reproduce the
surface temperature of modern Earth, which is close to the regime
simulated here. It may be that the cloud parameterizations are tuned to
compensate for differences in clear sky radiative transfer among the
models.

It is important to note, however, that AM2 and LMDG\_random are both
5--8 K colder than the other models (Table 2). Given that AM2 produced
very similar surface temperatures to CAM3 in the simulations without
clouds, we can attribute the difference in simulations with clouds to
differences in cloud parametrization: The lower temperature is due to
a stronger negative cloud radiative effect ($-$48~W~m$^{-2}$),
resulting in a higher planetary albedo. In fact, the cloud radiative
effect at the top of the model is correlated with surface temperature
in all the experiments, with a more negative cloud radiative effect
associated with a lower global-mean surface temperature (Table~2).
Similarly, the difference between LMDG\_max and LMDG\_random is
entirely due to clouds, and shows that in this configuration switching
from one extreme assumption on the cloud overlap to the other can have
a 8 K effect on the global-mean surface temperature.

Spatial patterns of cloud fraction as well as cloud water amount are similar 
among the models, but the magnitudes have very large differences. 
All models show broadly similar patterns of cloud
fraction, with pronounced Intertropical Convergence Zones (ITCZs) and
relatively low-level clouds at mid and high latitudes
(Fig.~\ref{fig:GStarClouds}(a--f)). The cloud fraction is generally higher
in AM2 than in the CAM models. This, in combination with potential
microphysical differences (such as cloud particle size), is likely why AM2 produces a 
more negative cloud radiative effect and lower surface temperatures, 
although its cloud water amount is the lowest among the models (Fig.~\ref{fig:GStarClouds}(g--l)). 
CAM3 is slightly cloudier than CAM4 and CAM4\_Wolf, as was found by
\cite{wolf2015evolution}, which likely causes its slightly lower
global-mean surface temperature. Cloud fraction in LMDG\_max 
is less than that in LMDG\_random while the cloud water amount is  
similar between the two versions of LMDG, so that clouds have 
a larger cooling effect in LMDG\_random, $-$55 versus $-$34~W~m$^{-2}$ 
(Table~2 and Fig.~\ref{fig:GStarClouds}(e--f, k--l)).

\subsection{M-star Planet GCM Comparison}
\label{sec:M-Star-Spectrum}

\begin{figure}[!htbp]  
\begin{center}
\includegraphics[angle=0, width=14cm]{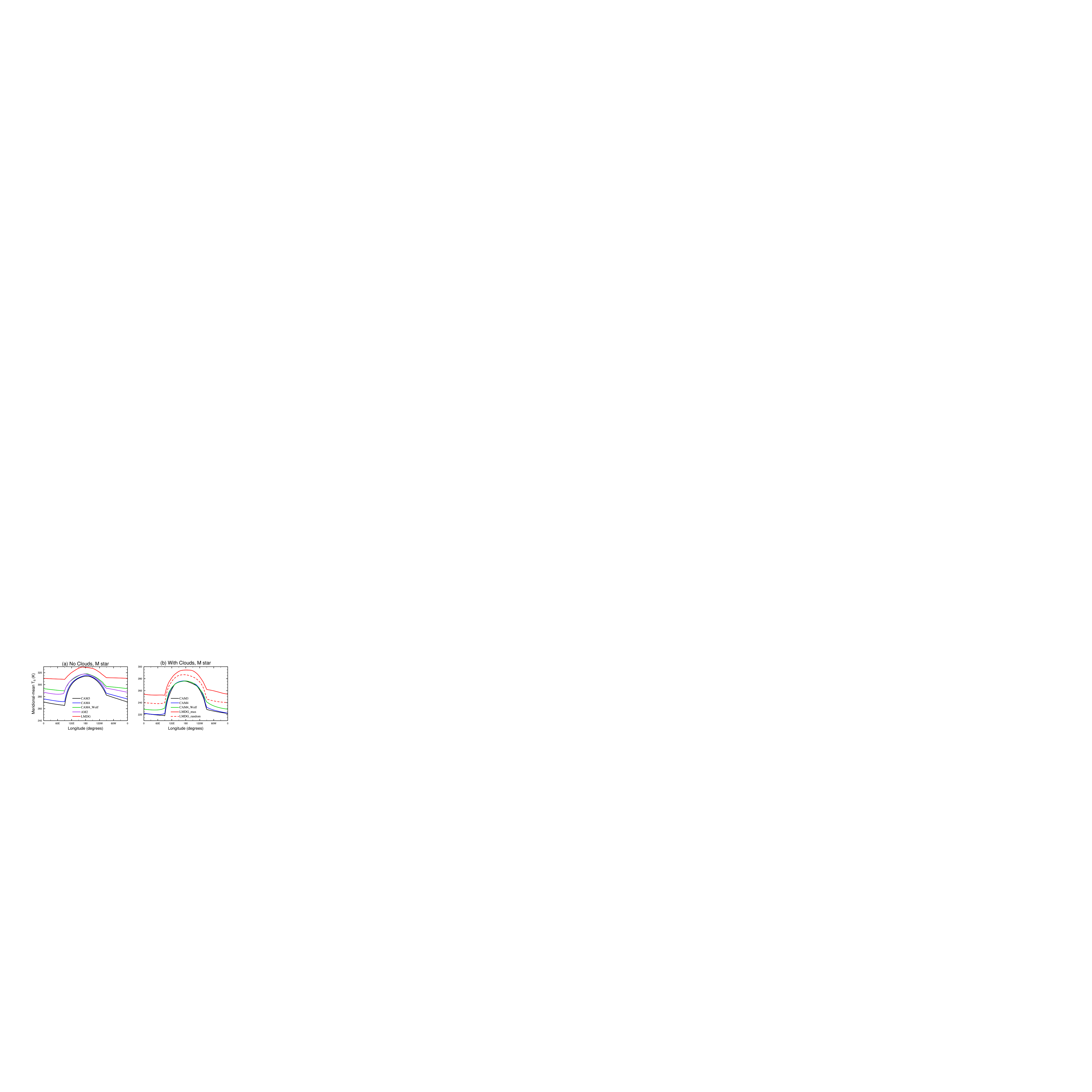}
\caption{M star surface temperature: Meridional (North-South) mean
  surface temperature as a function of longitude for GCMs both
  without clouds ((a), CAM3, CAM4, CAM4\_Wolf, AM2, and LMDG) and including clouds ((b), 
  CAM3, CAM4, CAM4\_Wolf, LMDG\_max, and LMDG\_random). The configuration
  assumes a tidally locked aqua-planet with an M star stellar spectrum
  and a stellar flux of 1,360~W~m$^{-2}$. The substellar point is at 180$^{\circ}$ longitude. 
  Note the different y axis ranges between (a) and (b).}
\label{fig:MStarComparison}
\end{center}
\end{figure}

\begin{figure}[!htbp]  
\begin{center}
\includegraphics[angle=0, width=14cm]{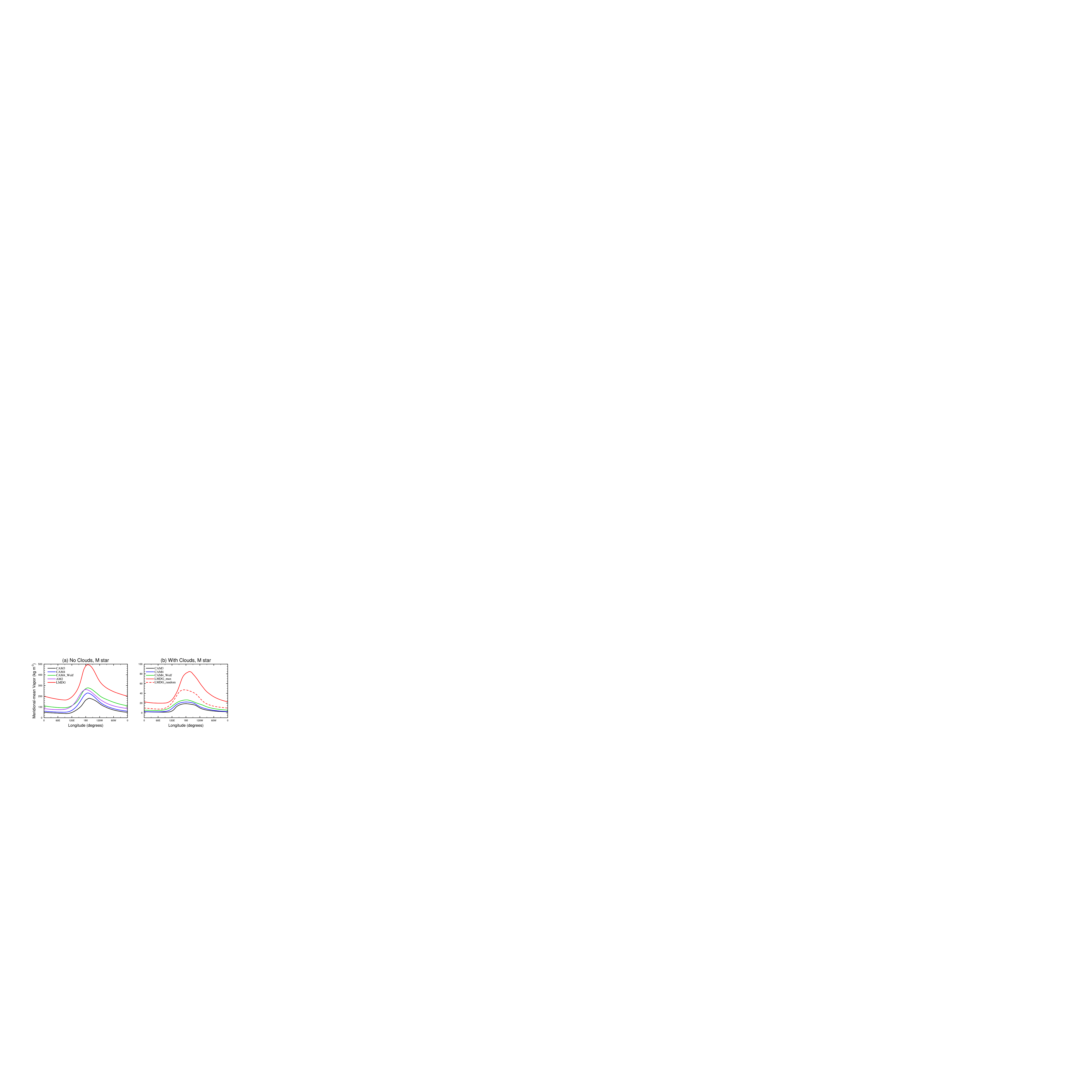}
\caption{M star water vapor content: Meridional (North-South) mean
  vertically integrated water vapor content in the atmosphere as a 
  function of longitude for GCMs both without clouds ((a), CAM3, CAM4, CAM4\_Wolf, 
  AM2, and LMDG) and including clouds ((b), CAM3, CAM4, CAM4\_Wolf, LMDG\_max, 
  and LMDG\_random). The configuration assumes a tidally locked 
  aqua-planet with an M star stellar spectrum and a stellar flux of 1,360~W~m$^{-2}$. 
  Note the different y axis ranges between (a) and (b).}
\label{fig:MStarH2O}
\end{center}
\end{figure}

When we ran the GCMs in tidally locked configuration with an M-star
spectrum, they produced larger differences than in the G-star case,
even without clouds (Fig.~\ref{fig:MStarComparison}(a)). CAM3 produced
the coolest climate, which is consistent with the fact that it has the
weakest greenhouse effect in 1D radiative-transfer mode \citep{Yang2016}. 
CAM4\_Wolf is warmer than CAM3, CAM4 and AM2; a
major cause of this is likely that the greenhouse effect of water
vapor in CAM4\_Wolf is the strongest among the four models
\citep{Yang2016}. The CAM models and AM2 show a range of behavior,
with differences in surface temperature among models particularly
pronounced on the night side. This is likely due to the large changes
in surface temperature possible if the strength of the night-side
temperature inversion changes. This effect is leveraged by increases
in the radiative time scale and atmospheric heat transport
\citep{koll-2016}, due to larger water vapor concentration in warmer
simulations. Parameterization of boundary layer turbulence could 
also influence the inversion strength and the night-side surface temperature. 
 Moreover, the water vapor feedback acts to amplify the
differences among models (Fig.~\ref{fig:MStarH2O}(a)). LMDG obtains
the highest global-mean surface temperature in the no-cloud
experiment, 14--28 K larger than other models (Table~3).  As in the
G-star, no-cloud experiment, although LMDG does not enter the runaway
greenhouse at this insolation (1,360 W\,m$^{-2}$) it is very close to
the runaway greenhouse. The absorbed stellar energy of the system in
this experiment is 326 W\,m$^{-2}$ in global mean (the planetary
albedo is 0.04)\footnote{Note that the runaway greenhouse limit
  depends on the orbital configuration of the planet.  For example, in
  the lower resolution LMG simulations described in
  Section~\ref{sec:explaining}, the absorbed stellar energy in the
  last converged solution is $\simeq$\,323~W~m$^{-2}$ in the tidally
  locked, no-cloud configuration, but it is $\simeq$\,315~W~m$^{-2}$
  in the rapidly rotating, no-cloud configuration
  (Fig.~\ref{fig:CAM3_LMDG}). The higher value in the tidally locked
  configuration is mainly due to the radiator fin effect of the
  permanent night side of a tidally locked orbit, which is relatively
  drier and can therefore emit longwave radiation to space more easily
  \citep{Pierrehumbert-1995:thermostats,yang2014-low}.}. Near or in
the runaway greenhouse state, outgoing longwave radiation at the top
of the atmosphere is insensitive to surface temperature and therefore
a large increase in the surface temperature is required to balance even a
very small increase in stellar radiation absorption
\citep{Pierrehumbert:2010-book}.

Note, a robust feature of these simulations is that, despite the lower
albedo, all models are cooler in the tidally locked setup that in the
rapidly rotating setup. This is mainly due to the radiator fin effect
of the permanent night side of a tidally locked orbit, which is
relatively drier and can therefore emit longwave radiation to space
more easily \citep{Pierrehumbert-1995:thermostats,yang2014-low}.

\begin{table}[!htbp]    
\label{tab:MstarTable}
\begin{center}
\caption{Global-mean climatic characteristics of the M-star spectrum GCM simulations. 
For the notes of different variables, please see Table~2. An insolation of 1,360~W\,m$^{-2}$ 
is close to the runaway greenhouse of LMDG when cloud radiative effects are turned off, 
explaining the high temperature of the model. The with-clouds case of AM2 met one 
unresolved problem, so that the experiment is not listed in the table.}
\begin{tabular}{llcccccccc}
\hline\hline
Simulations  & GCMs & $T_S$  &  $\alpha_p$      & SWCE & LWCE & NCE & Cld  & WVP & CWP \\
       & &  [K]  &   [0-1]   & [Wm$^{-2}$] & [Wm$^{-2}$] & [Wm$^{-2}$] & [\%]   & [kgm$^{-2}$]  & [kgm$^{-2}$]  \\
\hline
\multirow{5}{*}{No Clouds}    &  CAM3           & 288     & 0.04   &--- & ---&  ---& ---  & 84  &---\\
\multirow{5}{*}{}    &  CAM4           & 291       & 0.04    &--- & ---&---  &---& 106    & ---\\
\multirow{5}{*}{}   & CAM4\_Wolf  & 302      & 0.04     & ---& ---& --- &---& 155     & --- \\
\multirow{5}{*}{}   &  AM2              & 299      & 0.04   & ---& ---&---  &---   & 136   & ---\\
\multirow{5}{*}{}   &  LMDG           &  316      & 0.04  & ---& ---&---  &---   &  268  &--- \\
\hline
\multirow{5}{*}{With Clouds}  & CAM3             & 246         & 0.46     & --138  & 17 & --121 & 97 & 7  & 0.15 \\
\multirow{5}{*}{}  & CAM4             & 247         & 0.46     & --140  & 21 & --119 & 98 & 9   & 0.17 \\
\multirow{5}{*}{}  & CAM4\_Wolf   & 252         & 0.44  &  --131 & 19 & --112 & 98 & 13     & 0.19 \\
\multirow{5}{*}{}  & LMDG\_max       &  272   & 0.30    & --82 & 20  & --62   & 34 & 41   & 0.15 \\
\multirow{5}{*}{}  & LMDG\_random  &  262  & 0.38     & --108 & 22 & --86  & 81 & 22  & 0.11\\
\hline
\end{tabular}
\end{center}
\end{table}

\begin{figure}[!htbp]  
\begin{center}
\includegraphics[angle=0, width=16cm]{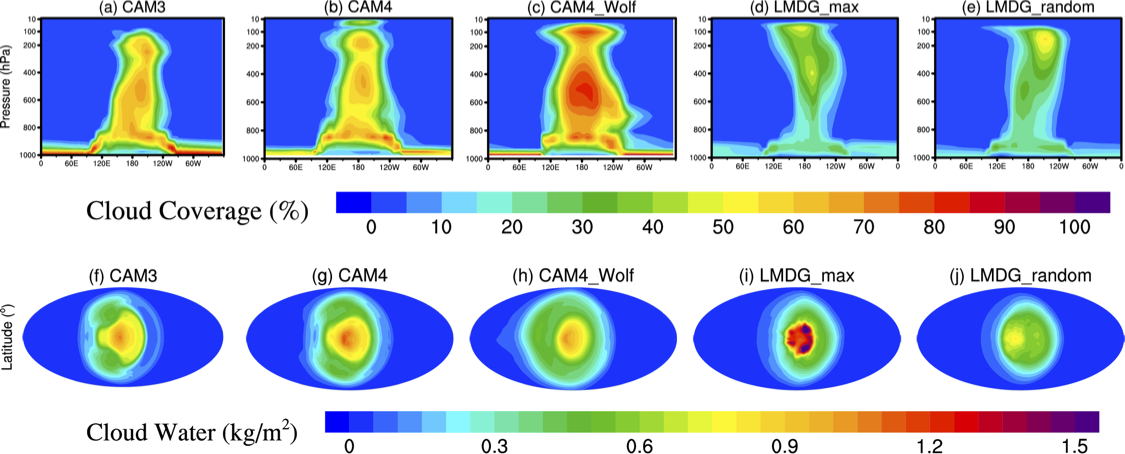}
\caption{M star clouds: Upper panels: Contour plots of meridional (North-South) mean
  cloud fraction as a function of longitude and pressure (vertical), 
  and lower panels: Vertically integrated cloud water amount (including both liquid 
  and ice phases), for GCMs 
  CAM3, CAM4, CAM4\_Wolf, LMDG\_max, and LMDG\_random. The configuration assumes a 
  tidally locked aqua-planet with an M star stellar spectrum and a stellar flux of
  1,360~W~m$^{-2}$.}
\label{fig:MStarClouds}
\end{center}
\end{figure}

With clouds included, the various versions of CAM yield surprisingly
similar surface temperatures (Fig.~\ref{fig:MStarComparison}(b)), 
especially given the variation in
M-star spectral energy distribution stellar absorption in 1D
radiative-transfer mode \citep{Yang2016}, although CAM4\_Wolf does have a
global-mean surface temperature about 5--6 K higher (Table~3). 
However, the global-mean surface temperature of LMDG 
is 10-26 K higher than those in CAM models. The
remarkable divergence among models emphasizes the fact that we should
not over-interpret the results of any single model when simulating
exoplanet climates.

The most striking feature of the GCM cloud simulation in the M-star,
tidally locked case (Fig.~\ref{fig:MStarClouds}) is that all models
confirm previous work
\citep{yang2013,way2015exploring,kopparapu2016inner,salameh2017role}
predicting deep convective clouds at the substellar point. We find
that LMDG has relatively low cloud fractions that are somewhat
weighted toward high altitude, optically thin clouds
(Fig.~\ref{fig:MStarClouds}(a--e)).
This contributes significantly to the fact that LMDG produces much
higher surface temperatures than the CAM models. The planetary albedos
in LMDG\_max and LMDG\_random are 0.30 and 0.38, respectively, about
0.16 and 0.08 lower than those in CAM3 (Table 3).  LMDG\_max has a
much lower cloud fraction but higher cloud water amount than
LMDG\_random (Fig.~\ref{fig:MStarClouds}(d--e, i--j)), such that 
LMDG\_max has a weaker net cloud radiative effect, $-$62 versus
$-86$~W~m$^{-2}$, and a warmer surface, 272 versus 262 K in global
mean.  Most models produce boundary layer clouds on the night side,
but these have very little radiative effect.  Again, the water vapor
feedback is important for enhancing differences among models
(Fig.~\ref{fig:MStarH2O}(b)).


\subsection{Explaining Differences Between CAM3 and LMDG}
\label{sec:explaining}

\begin{figure}[!htbp]
\begin{center}
\includegraphics[angle=0, width=14cm]{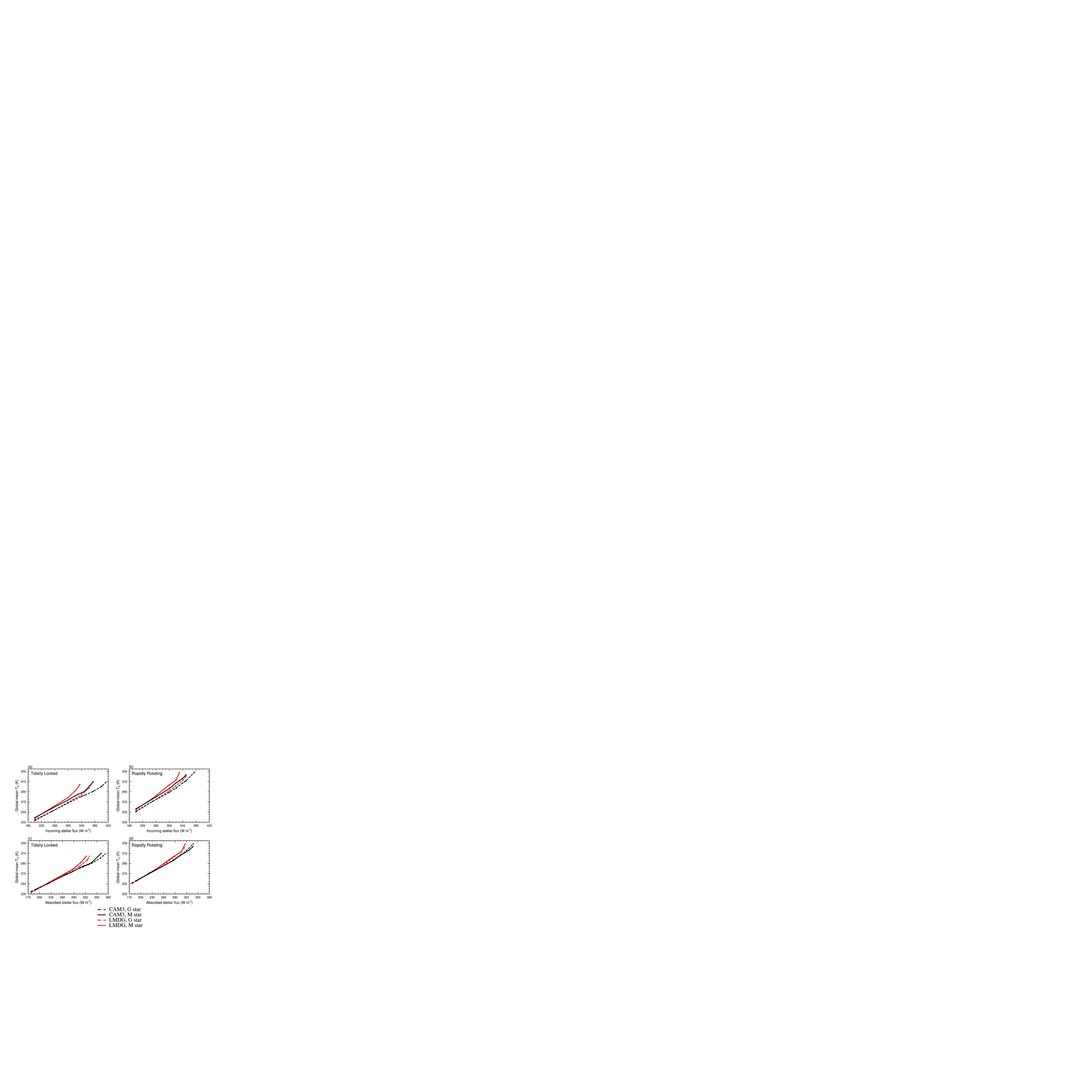}
\caption{Global-mean surface temperature as a function of global-mean
  incoming stellar flux (a, b) and absorbed stellar flux (c, d) for
  CAM3 (black lines) and LMDG (red lines) when the GCMs are run in a
  tidally locked aqua-planet configuration (a, c) and a rapidly
  rotating aqua-planet configuration (b, d) with a G star stellar
  spectrum (dashed lines) and with an M star stellar spectrum (solid
  lines). Clouds are turned off in all these simulations, and the
  surface albedo is 0.05 everywhere. Note that the maximum stellar
  fluxes at the substellar point of the tidally locked experiments in
  (a) and (c) are 1,650 W\,m$^{-2}$ (CAM3, G star), 1,500 W\,m$^{-2}$
  (CAM3, M star), 1,475 W\,m$^{-2}$ (LMDG, G star) and 1,340
  W\,m$^{-2}$ (LMDG, M star), and of the rapidly rotating experiments
  in (b) and (d) are 1,500 W\,m$^{-2}$ (CAM3, G star), 1,400
  W\,m$^{-2}$ (CAM3, M star), 1,400 W\,m$^{-2}$ (LMDG, G star) and
  1,320 W\,m$^{-2}$ (LMDG, M star). A further increase of the stellar
  flux in LMDG will push LMDG into a runaway greenhouse state or CAM3
  to blow up. Note: To speed up computations, we decreased the
  resolution of LMDG to 11.25$^{\circ}$$\times$5.625$^{\circ}$ in the
  simulations for this figure. Although this will not affect the
  trends discussed here, it may affect the exact
  location of the runaway greenhouse limit of the model.}
\label{fig:CAM3_LMDG}
\end{center}
\end{figure}

In order to investigate the mechanistic causes of differences among
GCMs in more detail, we performed additional simulations and analyses
of CAM3, a relatively cool GCM, and LMDG, a relatively warm
GCM. Fig.~\ref{fig:CAM3_LMDG} shows a comparison of the global mean
surface temperature in CAM3 and LMDG as a function of both incoming
stellar flux and absorbed stellar flux in both tidally locked and
rapidly rotating aqua-planet configurations forced by both G-star and
M-star spectral energy distributions. Clouds are set to zero in all of
these simulations. 

Both models are warmer for a given stellar flux when forced by an
M-star spectral energy distribution because water vapor absorbs longer
wavelengths of light better. Under the same incoming stellar flux, the
global-mean surface temperature in the rapidly rotating case is higher
than that in the tidally locked case. This is mainly due to the
cooling effect of the radiator fin of the permanent night side on
tidally locked planet
\citep{Pierrehumbert-1995:thermostats,yang2014-low}. When we plot as a
function of absorbed stellar flux, the difference between G-star and
M-star surface temperatures within a model is greatly reduced. Plotted
in this way though, a cold offset of CAM3 relative to LMDG appears,
and the offset becomes larger with increasing stellar flux. There are
several processes that may cause the differences between CAM3 and
LMDG, including radiative transfer, atmospheric dynamics, and
differences in the water vapor distribution due to differences in
convection parameterizations and dynamical processes. We will
investigate these below, focusing our attention on the tidally locked
case around an M star, where differences between the two models are
largest.

\subsubsection{Clear-sky Radiative Transfer}
\label{sec:RadiativeTransfer}
\begin{figure}[!htbp]
\begin{center}
\includegraphics[angle=0, width=14cm]{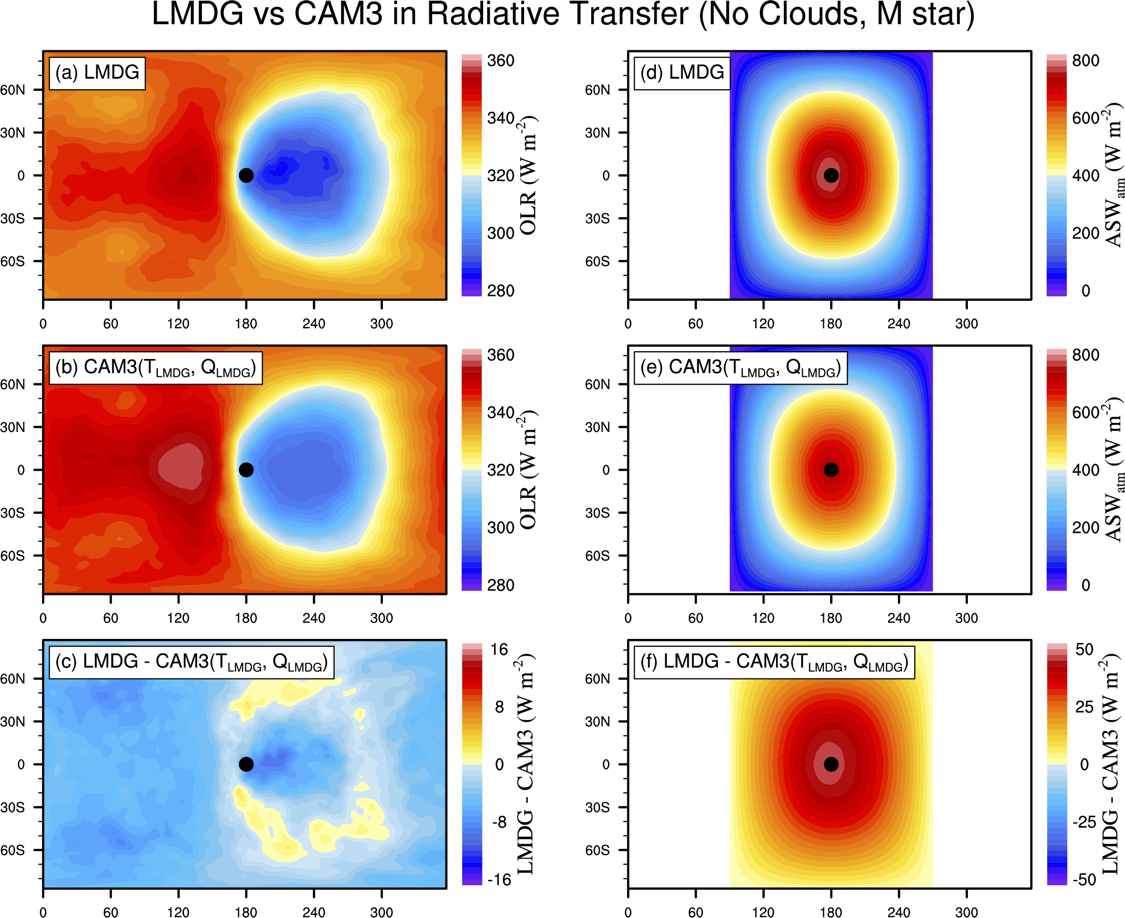}
\caption{Cloud-free radiative transfer in LMDG and CAM3: Outgoing
  longwave radiation (OLR) at the top of the model (left column) and
  absorbed shortwave flux by the atmosphere (ASW$_{atm}$) for LMDG and
  for CAM3 forced by the same temperature and water vapor profiles
  from LMDG. (a) OLR in LMDG, (b) OLR in CAM3, and (c) the difference:
  LMDG $-$ CAM3.  (d) ASW$_{atm}$ in LMDG, (e) ASW$_{atm}$ in CAM3,
  and (f) the difference: LMDG $-$ CAM3. The black dot is the
  substellar point.  The global-mean value is $-$3.8~W~m$^{-2}$ in (c)
  and 11.7~W~m$^{-2}$ in (f).}
\label{fig:LMDGvsCAM3_RT}
\end{center}
\end{figure}

When forced by the same 1D temperature and water vapor profiles, CAM3
absorbs less radiation in both infrared and visible wavelengths than
LMDG, i.e., CAM3 has a weaker greenhouse effect and smaller shortwave
energy absorption \citep{Yang2016}. These differences result from LMDG
using an updated HITRAN database, HITRAN2008 versus HITRAN2000 in CAM3
\citep{Yang2016}. HITRAN2008 has many more absorption lines and
stronger absorption cross sections in many wavelengths than HITRAN2000
(Supplementary Fig.~3 in \cite{Goldblatt:2013}). Moreover, there are
36 stellar spectrum intervals in LMDG and only 7 in CAM3
\citep{Yang2016}. The higher spectral resolution in LMDG allows it to
accurately resolve the individual absorption and window wavelengths
separately. We confirm these differences here by inputting the
simulated 3D temperature and water vapor profiles from LMDG into
CAM3's radiative transfer module (Fig.~\ref{fig:LMDGvsCAM3_RT}). We
find that the outgoing longwave radiation using CAM3's radiation is
higher than LMDG's by 3.8~W m$^{-2}$ in the global mean and the
absorbed shortwave radiation by the atmosphere using CAM3's radiation
is lower than LMDG's by 11.7~W m$^{-2}$ in the global mean. Both of these
 effects lead to a cooler climate in CAM3. Consistent with this finding,
\cite{kopparapu2017habitable} showed that accounting for these updated
line lists and continuum absorption coefficients 
reduces the stellar flux limit for the runaway greenhouse.

\subsubsection{Dry Dynamical Core}
\label{sec:DynamicalCore}

\begin{figure}[!htbp]
\begin{center}
\includegraphics[angle=0, width=16cm]{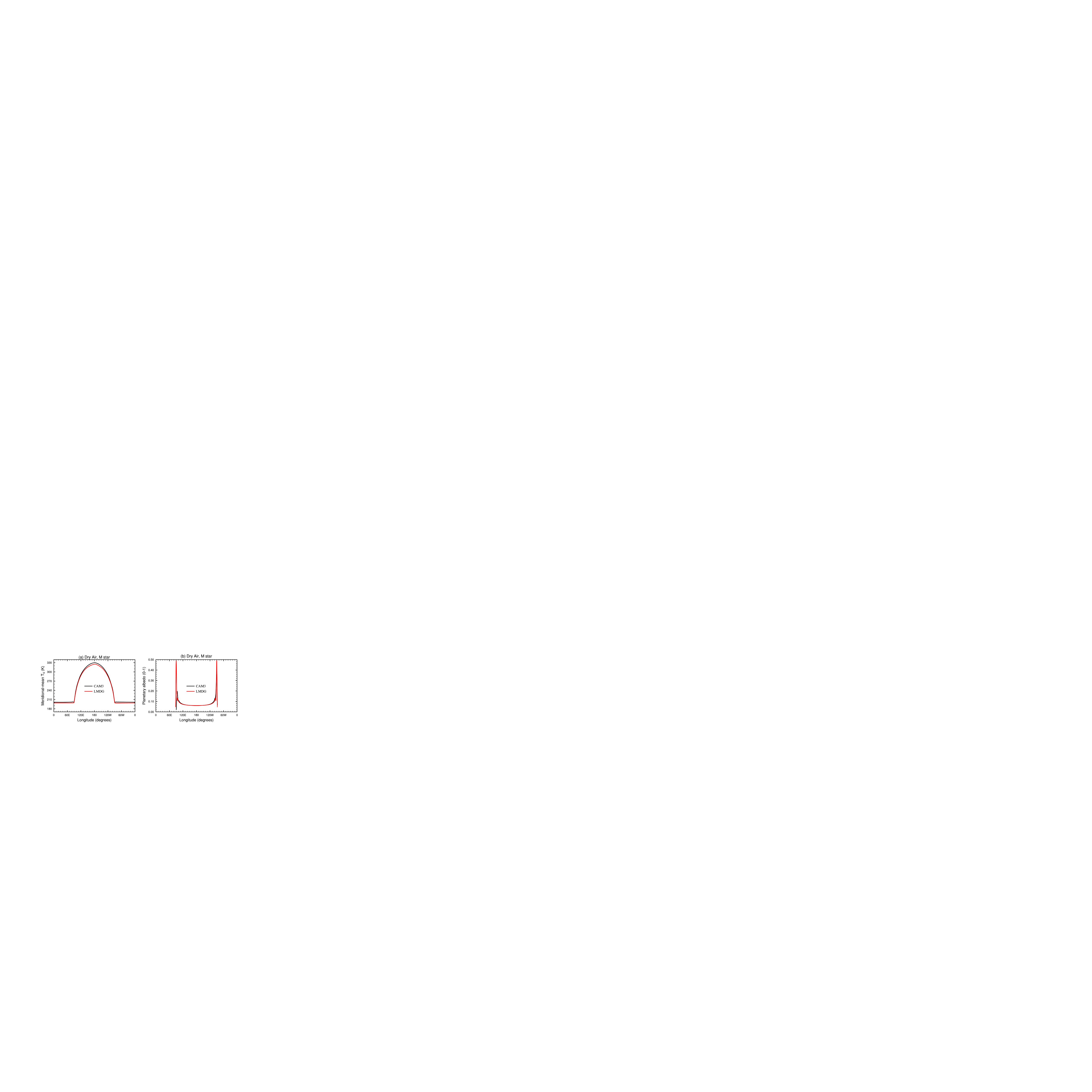}
\caption{Dry atmosphere simulations: Meridional (North-South) mean
  surface temperature (a) and planetary albedo (b) as a function of
  longitude for both CAM3 and LMDG assuming a nearly dry atmosphere on
  a tidally locked planet with a stellar flux of 1,360~W~m$^{-2}$.  The
  water vapor mixing ratio is set to 10$^{-6}$ everywhere, which is
  the minimum vapor concentration allowed in the radiative transfer
  correlated-$K$ tables of LMDG.}
\label{fig:DryComparison}
\end{center}
\end{figure}

To test for dynamical differences between the two models, we performed
tidally locked simulations with an M-star spectral energy
distribution, no clouds, and with atmospheric water vapor mixing ratio
set to 10$^{-6}$ everywhere, which is the minimum vapor concentration
covered by the radiative transfer correlated-$K$ tables in LMDG. The
surface temperature simulation was nearly identical between the two
models in this case (Fig.~\ref{fig:DryComparison}), in fact, LMDG was
actually slightly cooler than CAM3. This is in striking contrast to
the same simulation performed with water vapor and clouds, where LMDG
produced a climate much warmer than CAM3
(Table 3). This test shows that dry dynamics
alone do not contribute significantly to differences in the simulation
of climate in LMDG and CAM3.

\subsubsection{Relative Humidity}
\label{sec:RelativeHumidity}

\begin{figure}[!htbp]
\begin{center}
\includegraphics[angle=0, width=16.5cm]{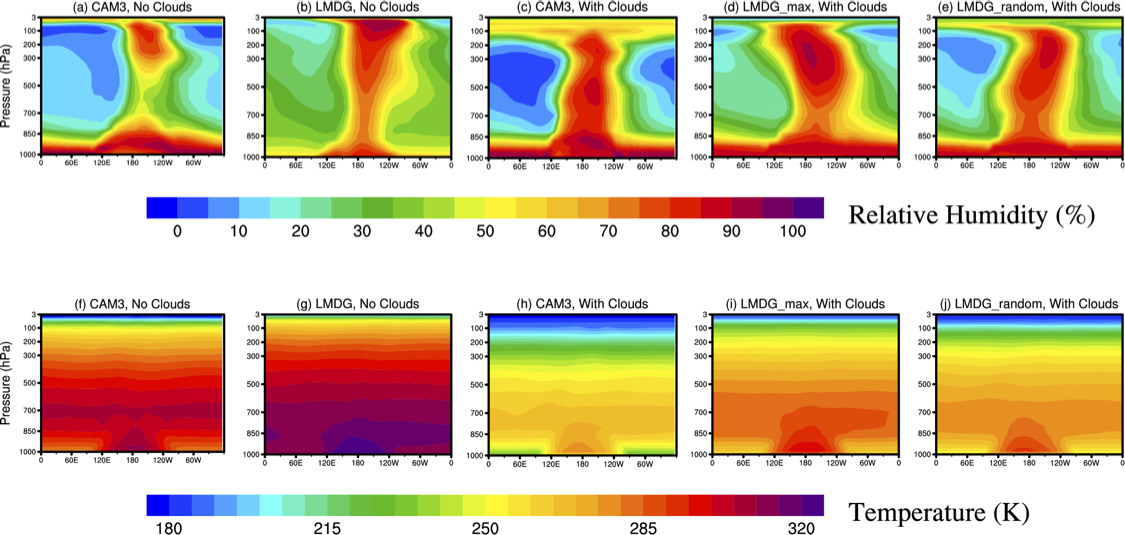}
\caption{M star relative humidity and air temperature: Contour plots of meridional
  (North-South) mean relative humidity (upper panels) 
  and air temperature (lower panels) as a function of longitude 
  and pressure (vertical) for CAM3 and LMDG. The 
  simulations are for a tidally locked aqua-planet with an M star
  stellar spectrum and a stellar flux of 1,360~W~m$^{-2}$. Simulations
  both without clouds (a, b, f, g) and with clouds (c--e, h--j) are plotted.}
\label{fig:rhComparison}
\end{center}
\end{figure}

We next turn our attention to the simulation of relative humidity ($RH$),
which is defined as the percentage of water vapor 
mixing ratio\footnote{An alternative definition is ratio of the vapour pressure to the saturation vapor pressure (such as \cite{Vallis2017book}). The American Meteorological Society (AMS) uses the vapor pressure to define $RH$ while the World Meteorological Organization (WMO) used the mixing ratio to define $RH$ (glossary.ametsoc.org). The differences between these two definitions are very small when the water vapor is dilute.} relative to the saturation water vapor mixing ratio \citep{WallaceandHobbs2006book,Abbot2018}
and is a critical term for inferring habitability
\citep{Pierrehumbert-1995:thermostats,leconte2013increased,PierrehumbertandDing2016} 
that can be affected by both radiative transfer and atmospheric dynamics.
Relative humidity is higher in LMDG than in CAM3 at high altitude
around the planet, both with and without clouds (Fig.~\ref{fig:rhComparison}). High-altitude water
vapor is particular important because it increases the optical
thickness in a cold region of the atmosphere, causing strong
greenhouse warming. The fact that the high-altitude relative humidity
is much higher in LMDG than in CAM3 is likely one of the causes of the
much higher surface temperature in LMDG.


Atmospheric relative humidity is determined by many processes,
including large-scale atmospheric circulation, eddies, and small-scale
processes such as convection, entrainment, detrainment, re-evaporation
of rain droplets, and diffusion
\citep{pierrehumbert2007relative,Sherwood:2010RG,Wrightetal2010RH}.
Here, we can identify two factors that likely make LMDG moister. The first
is that LMDG uses a forced convective adjustment
\citep{Manabe-Wetherald-1967:thermal} to calculate the atmospheric
lapse rate, whereas CAM3 determines the atmospheric lapse rate
prognostically based on complex moist processes
\citep{wolf2015evolution}.  This difference can have a big impact on
moisture distributions, as \citet{wolf2015evolution} found when they
compared LMDG and CAM4\_Wolf, which determines the lapse rate in a
similar way to CAM3.  Specifically, they showed that in hot climates,
even under the same global-mean surface temperature, LMDG's upper
atmosphere is always much moister than CAM4\_Wolf's. Next we suggest
a second reason LMDG may be moister: differences in shortwave
absorption.

\begin{figure}[!htbp]
\begin{center}
\includegraphics[angle=0, width=14cm]{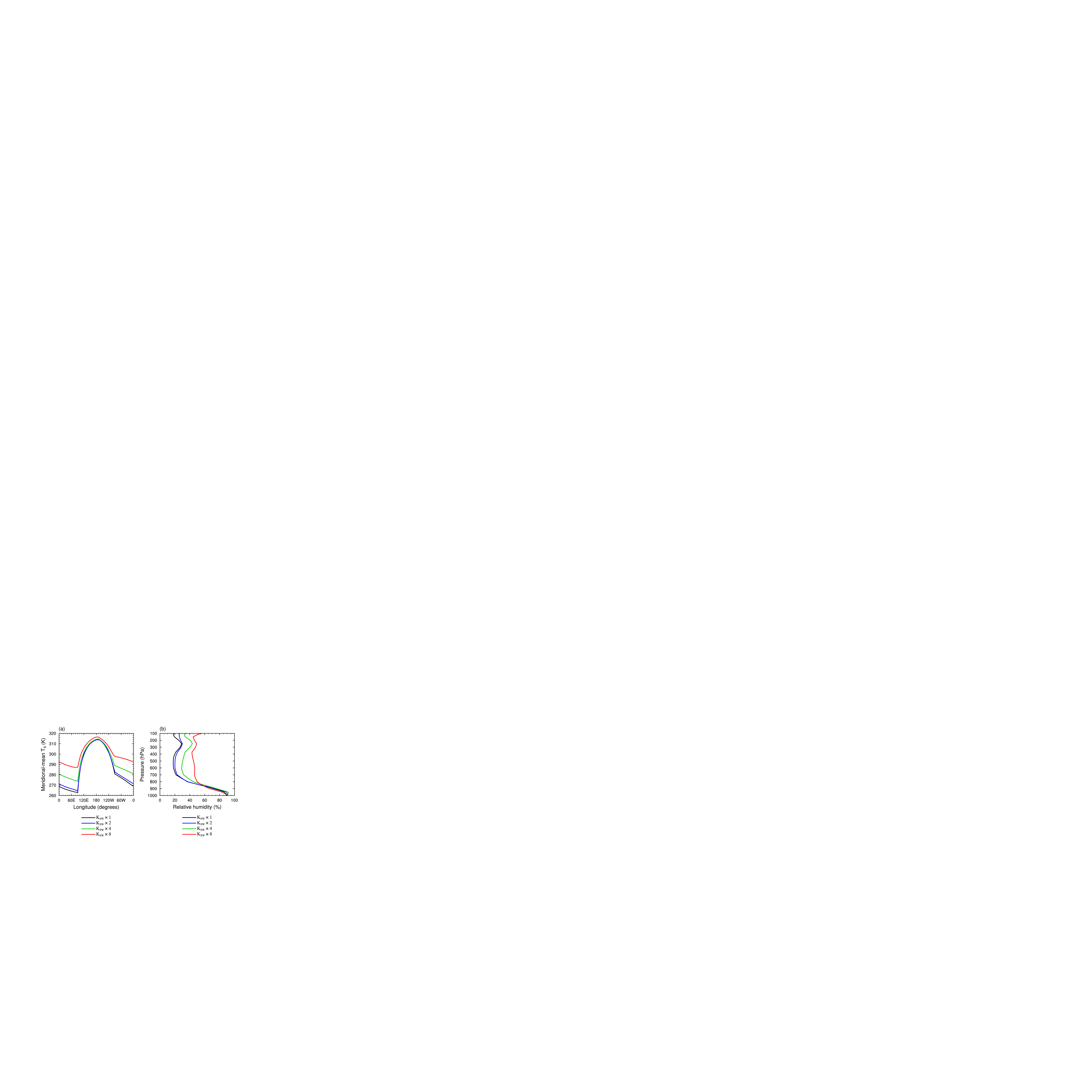}
\caption{Varying the shortwave absorption coefficient of water vapor
  ($K_{H_2O}$) in CAM3: Meridional (North-South) mean surface
  temperature as a function of longitude (left) and global mean
  vertical profiles of relative humidity (right) for the simulations
  with $K_{H_2O}$ increased by a multiple of 1, 2, 4, or 8. The
  simulations are for a tidally locked aqua-planet with an M star
  stellar spectrum, a stellar flux of 1,292~W~m$^{-2}$, and without
  clouds. The surface albedo is set to zero everywhere, so that the
  planetary albedo is close to zero in all of these cases (not
  shown). Note that the 4 and 8 times absorption coefficients are
  unrealistic. Increasing the absorption coefficients by about twice
  in CAM3 is able to approximately match the shortwave heating rates
  in LMDG (see Fig.~\ref{fig:SWH_cam3LMDG} below).}
\label{fig:kh2o_cam3}
\end{center}
\end{figure}

\begin{figure}[!htbp]
\begin{center}
\includegraphics[angle=0, width=16cm]{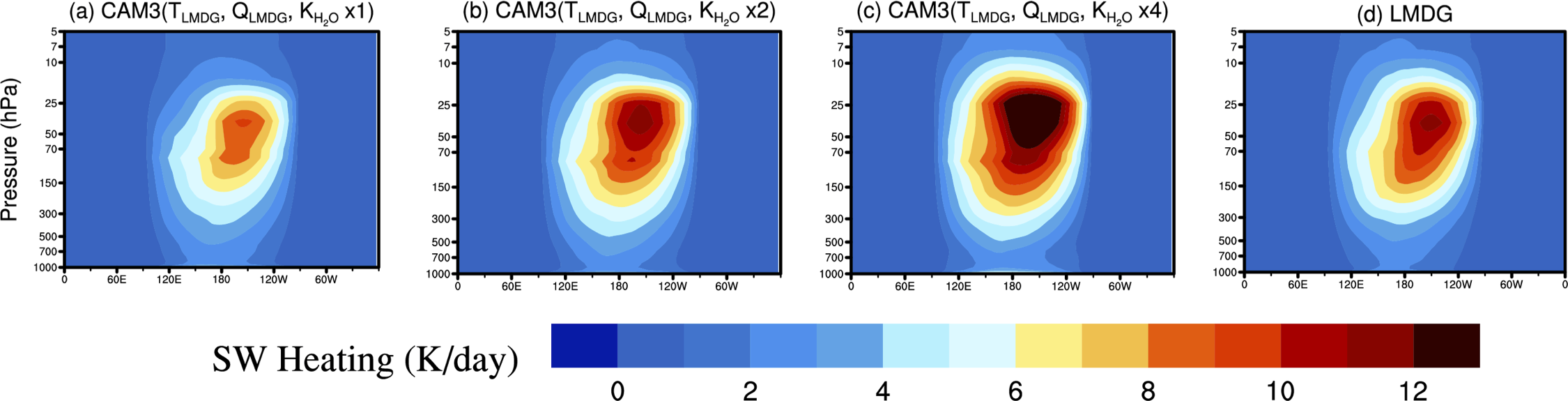}
\caption{Meridional (North-South) mean shortwave heating rate: CAM3
  versus LMDG.  (a) Using CAM3's default shortwave absorption
  coefficient of water vapor ($K_{H_2O}$), (b) doubling the values of
  $K_{H_2O}$, (c) quadrupling the values of $K_{H_2O}$, and (d) LMDG. 
  In the calculations, CAM3's radiative transfer
  module is forced by temperature and water vapor profiles
  from LMDG.  The calculations are for a tidally locked aqua-planet
  with an M star stellar spectrum, a stellar flux of 1,360~W~m$^{-2}$,
  and without cloud radiative effects.}
\label{fig:SWH_cam3LMDG}
\end{center}
\end{figure}

\begin{figure}[!htbp]
\begin{center}
\includegraphics[angle=0, width=16cm]{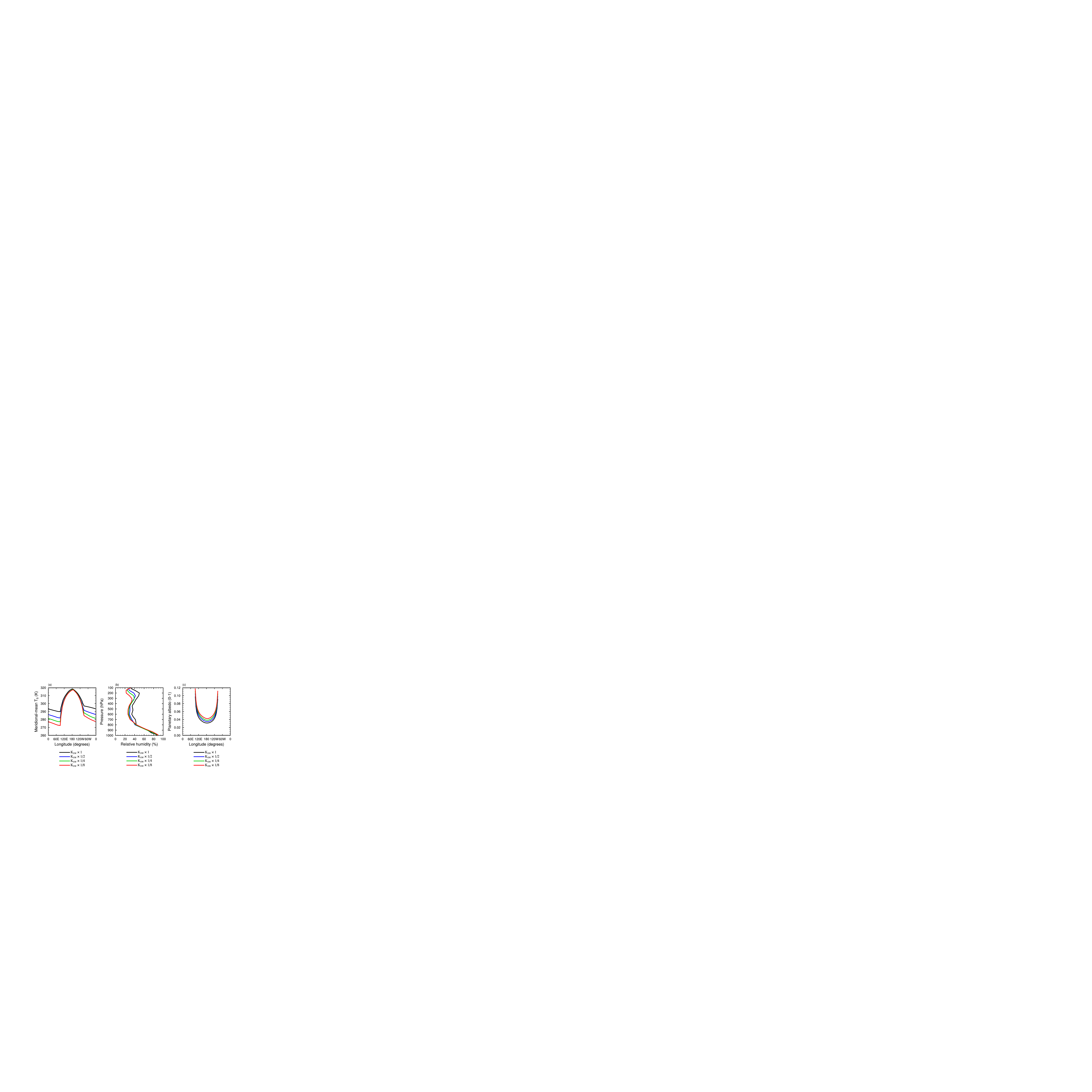}
\caption{Varying the shortwave absorption coefficient of water vapor
  ($K_{H_2O}$) in CAM4\_Wolf: Meridional (North-South) mean surface
  temperature (a) and planetary albedo (c) as a function of longitude,
  and global-mean vertical profiles of relative humidity (b) for the
  simulations with $K_{H_2O}$ decreased by a multiple of 1, 1/2, 1/4,
  or 1/8. The simulations are for a tidally locked aqua-planet with an
  M star stellar spectrum, a stellar flux of 1,360~W~m$^{-2}$ and
  without clouds.  The global-mean surface temperatures are 302, 297,
  295 and 292 K, and the planetary albedos are 0.044, 0.049, 0.053 and
  0.057, respectively.  The surface albedo is 0.05 everywhere. The
  tiny changes in planetary albedo are not enough to explain the
  changes in surface temperature.}
\label{fig:kh2o_cam4WT}
\end{center}
\end{figure}

A major difference between CAM3 and LMDG is that absorption of 
stellar radiation by water vapor is significantly higher in LMDG than
in CAM3 (section~\ref{sec:RadiativeTransfer}). When we artificially
increased the shortwave water vapor absorption coefficient in CAM3, we
found that this significantly increased the high-altitude relative
humidity and surface temperature (Fig.~\ref{fig:kh2o_cam3}).  When we
doubled the absorption coefficient by water vapor, the shortwave heating
rate of the atmosphere in CAM3 is close to that in LMDG
(Fig.~\ref{fig:SWH_cam3LMDG}), the global-mean surface temperature
increases by 0.9 K, and the night-side surface temperature increases by
2.5 K\footnote{Although seemingly small, this effect would be further
  amplified by the strong positive water vapor radiative feedback if
  another source of heating---an increase in longwave absorption for
  example---were to be added.}.  In CAM4\_Wolf, we find the same
phenomenon: When the shortwave water vapor absorption coefficient is
decreased, the high-altitude relative humidity decreases and the
surface cools (Fig.~\ref{fig:kh2o_cam4WT}). To understand this, we
built a last saturation model for water vapor
\citep{pierrehumbert2007relative} in which we trace air parcels and
approximate their specific humidity as its value the last time the
parcel was saturated (see
Appendix~\ref{sec:LastSaturationModel}). Model resolution and
numerical diffusion limit the accuracy of this method, but we are able
to broadly reproduce the high-altitude relative humidity.

\begin{figure}[!htbp]
\begin{center}
\includegraphics[angle=0, width=7cm]{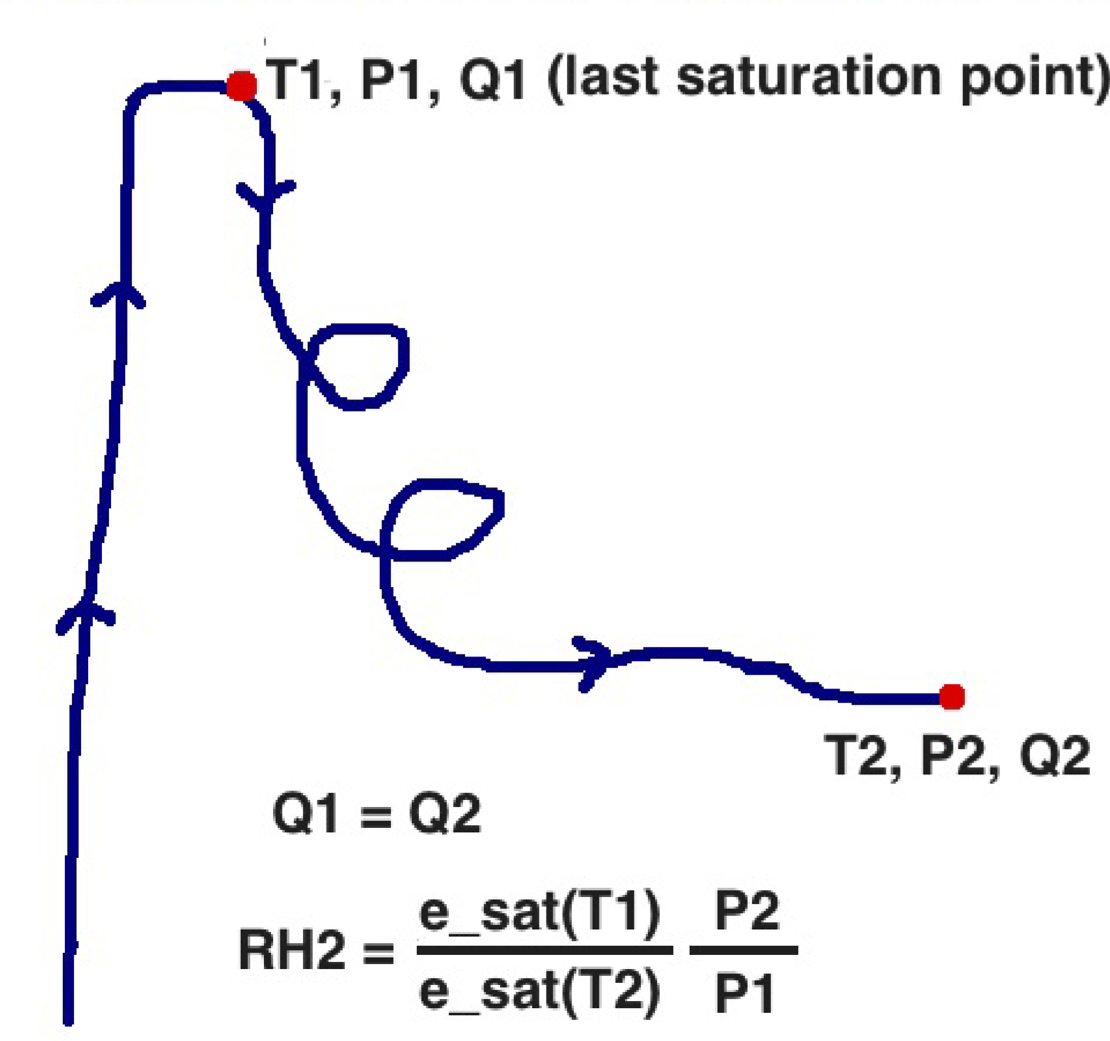}
\caption{Schematic illustration of the last saturation model for an
  air parcel. Specific humidity is conserved after the time of last
  saturation.  $T$ represents the air parcel's temperature, $P$ its
  pressure, and $Q$ its specific humidity.}
\label{fig:LastSatModel}
\end{center}
\end{figure}

\begin{figure}[!htbp]
\begin{center}
\includegraphics[angle=0, width=14cm]{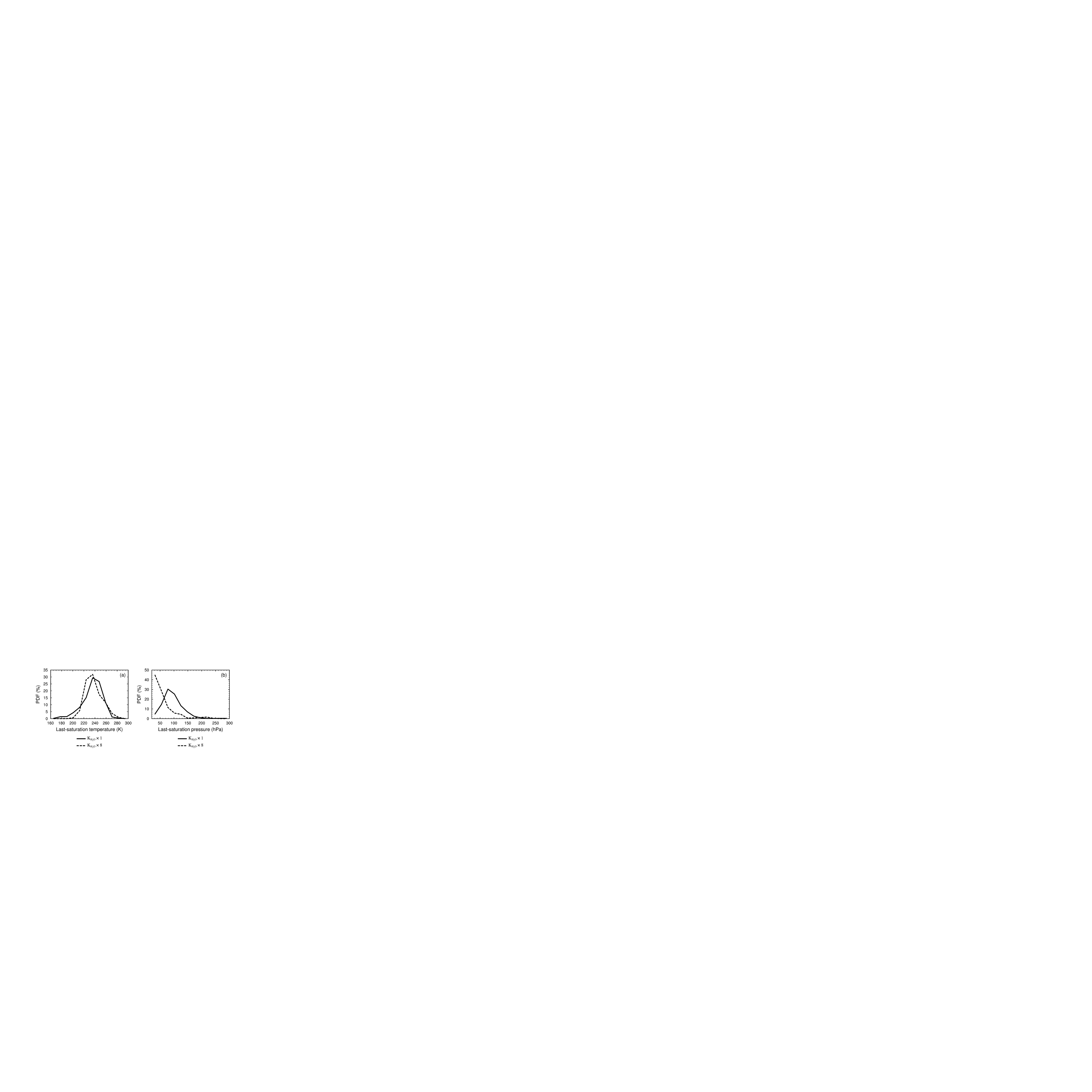}
\caption{Probability distribution functions (PDFs) of the last saturation temperature 
((a), $T_1$ in Fig.~\ref{fig:LastSatModel}) and pressure ((b), $P_1$ in 
Fig.~\ref{fig:LastSatModel}) for CAM3 run with normal 
  shortwave water vapor absorption coefficient $K_{H_2O}$ (solid) 
  and with eight times $K_{H_2O}$ (dashed).}
\label{fig:LastSatPDF}
\end{center}
\end{figure}

As air parcels rise in convection in the substellar region, they tend
to experience detrainment and convective outflow at some pressure
associated with an anvil cloud. This will mark the point of last
saturation, as the air is subsequently advected away from the
substellar point and descends to higher pressures as it cools
radiatively and heats adiabatically. As shown 
in Fig.~\ref{fig:LastSatModel}, for a given pressure in the 
descending region ($P_2$), the relative humidity ($RH_2$) 
is determined by the temperature and the air pressure at 
the last saturation point ($T_1$ and $P_1$), and can be approximately written as,
\begin{equation}
RH_2 = 100  \frac{e(T_2)}{e_{sat}(T_2)}   \frac{P_2 - e_{sat}(T_2)}{P_2 - e(T_2)} 
\approx 100 \frac{e(T_2)}{e_{sat}(T_2)}  \approx 100 \left( \frac{e_{sat}(T_1)}{e_{sat}(T_2)} \right) \times \left( \frac{P_2}{P_1} \right), 
\end{equation}  
where $e_{sat}$ is the saturation vapor pressure, 
$e$ is the vapor pressure, and we have assumed the
vapor pressure is much less than the air pressure at and after the last saturation. 
It has previously been shown that the Fixed Anvil Temperature hypothesis
\citep{Hartmann-Larson-2002:important,Kuang-Hartmann-2007:FAT,Thompson-Bony-Li:FAT}  
holds fairly well for convection near the substellar point of tidally locked simulations in
CAM3 \citep{yang2014-low}. This implies that the temperature of the
point of last saturation ($T_1$) and the corresponding saturation
vapor pressure ($e_{sat}(T_1)$) should stay roughly constant\footnote{The FAT hypothesis denotes that the temperature at the detrainment level of tropical convective anvil clouds is nearly constant during climate change. The underlying mechanism is that energy balance in the tropical troposphere is primarily between convective heating by latent heat release in regions of deep convection and radiative cooling by longwave emission to space in clear-sky regions with large-scale subsidence. Because of this, the detrainment level of anvil clouds should be located at the altitude where the clear-sky radiative cooling diminishes rapidly. The clear-sky radiative cooling rate in the upper troposphere is primarily determined by water vapor emission. The temperature at which the saturation water vapor pressure becomes small enough that water vapor emission is ineffective is constrained by local air temperature because of the Clausius--Clapeyron relationship. Therefore, the temperature at the top of anvil clouds should be nearly independent of surface temperature. For the simulations without clouds in our study, we turn off the cloud radiative effects but cloud formation, latent heat release, precipitation, and clear-sky radiative transfer still exist, so that the FAT hypothesis works in our simulations.} as we
increase the shortwave water vapor absorption coefficient, but the
pressure ($P_1$) should decrease due to surface warming and an
increase in the altitude of the anvil cloud. This is exactly what we
see in CAM3 (Fig.~\ref{fig:LastSatPDF}).  Since the temperature at
last saturation ($T_1$) does not change much as the shortwave water
vapor absorption coefficient is increased, the saturation vapor
pressure ($e_{sat}(T_1)$) does not change much. But the air pressure ($P_1$)
decreases, so the specific humidity at last saturation must
increase. Another way to explain this is that there is the same amount
of water vapor (same temperature), but much less dry air (lower
pressure), so the water vapor specific humidity increases. This means
the relative humidity will be higher all along the air parcel's
subsequent trajectory, and explains why increasing the shortwave water
vapor absorption coefficient increases the high-altitude relative
humidity throughout the planet.

In addition to the shortwave water vapor absorption coefficient, we
have tested the sensitivity of CAM3 to a number of other
parameters. These include: numerical momentum diffusion near the top
of the model (0.1 or 100 times of the default value), surface momentum
transfer coefficient (0.1 or 10 times of the default value), sensible
and latent heat exchange coefficients (0.5 or 2 times of the default
value), deep and shallow convection relaxation timescales (from 0.1 hr
to 16 hrs), deep and shallow convection precipitation efficiencies
(0.1 or 10 times of the default value), deep convection downdraft mass
flux factor (from 0 to 0.7), relative humidity limit for large-scale
condensation (from 50\,\% to 99.9\,\%), convective and large-scale
precipitation evaporation efficiencies (0.1 or 10 times of the default
value), and the critical Richardson number for planetary boundary
mixing (from 0.1 to 1). In all of these tests, the global-mean surface
temperature in the cloud-free, M-star, tidally locked configuration is
within the range of 286--294 K, indicating that varying one single
parameter can induce a global-mean surface temperature difference
within 8 K. More work would be required to test the effect of varying two
or more parameters simultaneously.

\subsubsection{Summary}
\label{sec:summary}

To summarize, there are a number of differences between LMDG and CAM3
that lead to CAM3 simulating a much colder climate for tidally locked,
M-star planets. Differences in the models' radiative schemes lead to
LMDG absorbing more stellar radiation and emitting less planetary
radiation to space. The interplay between atmospheric dynamics and
cloud parameterization leads to a higher cloud fraction and cloud
optical thickness at the substellar point in CAM3, causing significant
cooling. Finally, moist processes and the interplay between the absorption of stellar
radiation and atmospheric dynamics leads to higher relative humidity
at high altitudes in LMDG, and therefore lower planetary thermal 
radiation to space, causing significant warming of LMDG.

\section{Conclusion and Discussion}
\label{sec:conclusiondiscussion}

In this paper we have performed an intercomparison of the 3D global
climate models CAM3, CAM4, CAM4\_Wolf, AM2, and LMDG 
both with and without clouds. Our conclusions are as follows:

\begin{enumerate}
\item When run with clouds for rapidly rotating planets receiving a
  G-star spectral energy distribution and a stellar flux of
  1,360~W~m$^{-2}$, the models produce global mean surface temperatures
  within 8~K. Small differences in cloud parameterization
  assumptions can lead to this level of variation, as shown by
  the LMD\_max and LMD\_random simulations.

\item When run with clouds for tidally locked planets receiving an
  M-star spectral energy distribution and a stellar flux of
  1,360~W~m$^{-2}$, the GCM's behavior is much more divergent (up to 26
  K in global-mean surface temperature). LMDG\_max is much warmer than
  the other models. Clouds are an important part of the reason for
  this behavior, but large differences among the models with clouds
  set to zero demonstrate that model divergence is also due to
  clear-sky radiative effects of water vapor, as well as the interaction of radiation
  with atmospheric dynamics.

\item We implemented a last saturation model for relative humidity and
  used it to show that a larger shortwave water vapor absorption in
  GCM leads not only to direct warming by decreasing the planetary
  albedo, but also to indirect warming by increasing the high-altitude
  relative humidity around the planet and therefore decreasing
  planetary radiation emitted to space (increasing the greenhouse
  effect).

\end{enumerate}

\begin{figure}[!htbp]
\begin{center}
\includegraphics[angle=0, width=14cm]{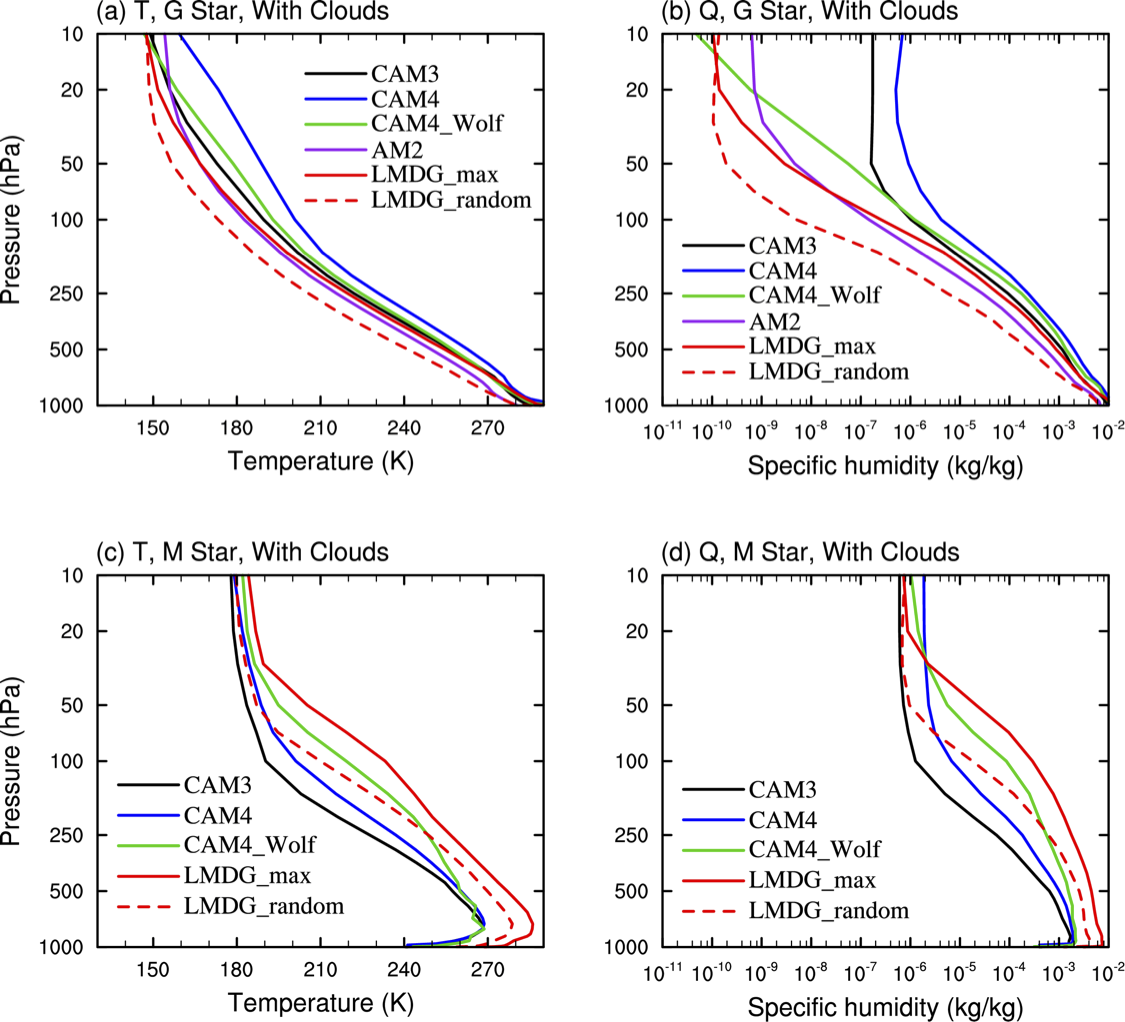}
\caption{Global-mean temperature profiles (a \& c) and specific humidity profiles (b \& d). 
  (a \& b) The simulations assume
  a rapidly rotating aqua-planet with a G star stellar spectrum, a
  stellar flux of 1,360~W~m$^{-2}$, and clouds. (c \& d) The simulations 
  assume a tidally locked aqua-planet with an M star stellar spectrum, 
  a stellar flux of 1,360~W~m$^{-2}$, and clouds. No ozone is included in 
  all of the simulations.}
\label{fig:TandQ}
\end{center}
\end{figure}

Besides of the differences in surface temperature, air temperature, and relative humidity 
between the models, there are also significant differences in stratospheric water vapor 
concentration (Fig.~\ref{fig:TandQ}), which influences the onset of the moist greenhouse 
state and the location of the inner edge of the habitable zone. From this figure, one could find 
that the stratospheric water vapor concentration is not directly connected to surface 
temperature, and it is more directly determined by air temperatures at high altitudes. 
The surface temperature difference among the models in the M-star, tidally locked 
experiments is larger than in the G-star, rapidly rotating experiments, but the stratospheric 
water vapor difference above 30 hPa is smaller in the former group of experiments. 
The strength of stratospheric circulation, such as the Brewer-Dobson circulation on 
Earth, can also influence the stratospheric water vapor  \citep{Holtonetal1995,Danielsen1993,FueglistalerandHaynes2005,RompsandKuang2009}. 
Future work is required to analyze the differences in stratospheric circulation and troposphere--stratosphere water vapor exchange between the models.

Our results are useful in explaining the differences between models 
those have been employed to examine the location of the inner edge of the habitable zone, such 
as why LMDG enters a runaway greenhouse state in a lower stellar radiation 
than that in the CAM models \citep{Leconte:2013gv,yang2013,wolf2015evolution}. 
Moreover, our results suggest that future work in developing exoplanet 
climate models should focus on improving the radiative transfer of 
water vapor in both longwave and shortwave and updating the cloud 
parameterization. In the present GCMs, the accuracy in shortwave radiative transfer 
is lower than that in longwave radiative transfer. 
Before direct atmospheric observations of exoplanets, 
laboratory cloud experiments and high-resolution 
cloud resolving models could be employed to investigate the clouds 
under different planetary parameters and the results could be used to 
improve the cloud parameterization in GCMs.
When interpreting the differences and similarities among the models
considered here, it is important to emphasize that agreement among
some or most of the GCMs does not imply that the climates they
simulate are correct. This is particularly true for the CAM models,
which share a similar heritage, and therefore share many similar or
identical subroutines. Determining which GCMs are the most accurate
requires detailed comparison with observations from Earth, other solar
system planets, and eventually observations of exoplanets. 
We should also remember that a GCM might 
perform better in one context and worse in another.

Since we performed the simulations for this intercomparison, 
three new planetary GCMs have been developed: 
Resolving Orbital and Climate Keys of Earth and Extraterrestrial 
Environments with Dynamics (ROCKE-3D) \citep{way2017resolving}, the 
Met Office Unified Model (UM) \citep{boutle2017exploring}, and Isca 
\citep{Vallisetal2018} as well as others.  Readers should be aware of these models 
and future intercomparison efforts should include them.


\acknowledgments \textbf{Acknowledgments:} We acknowledge support from NASA grant
number NNX16AR85G, which is part of the ``Habitable Worlds'' program
and from the NASA Astrobiology Institutes Virtual Planetary
Laboratory, which is supported by NASA under cooperative agreement
NNH05ZDA001C. J.Y. acknowledges support from the National Science
Foundation of China (NSFC) grants 41861124002, 41675071, 41606060, and
41761144072. J.L. acknowledges that this project has received funding 
from the European Research Council (ERC) under the European Union’s 
Horizon 2020 research and innovation programme (grant agreement 
number 679030/WHIPLASH). We are grateful to Junyan Xiong for his help in
drawing Figures 13 and 16. This work was completed in part with resources
provided by the University of Chicago Research Computing Center.

\clearpage

\appendix

\section{Last Saturation Model}
\label{sec:LastSaturationModel}

\begin{figure}
\begin{center}
\includegraphics[angle=0, width=16cm]{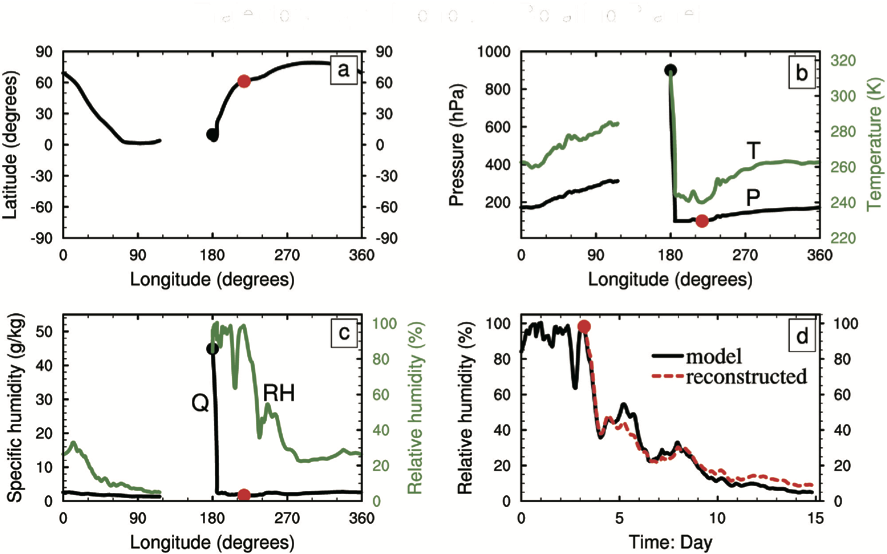}
\caption{This is an example of the trajectory of an air parcel that we
  trace using the last saturation model. The black dot in all
  panels shows where we begin to trace the parcel and the red dot 
  shows where it last reaches saturation. Panel (a) shows the latitude
  and longitude of the parcel as it rises near the substellar point
  and is advected away from the substellar point at altitude. Panel
  (b) shows the air pressure and temperature of the parcel as a
  function of longitude as it makes its voyage. Last saturation is
  achieved at the coldest air temperature reached. Panel (c) shows the
  specific and relative humidities of the parcel as a function of
  longitude. Panel (d) shows the GCM relative humidity (black line)
  and the relative humidity reconstructed from the last saturation
  model (red line) as a function of time. The red dot has a relative
  humidity of 90\,\% rather than 100\,\% (same as 
  \cite{Wrightetal2010RH}); this is because of the large grid size of
  the GCM (about 300 km) which means much of the air would saturated when
  the grid-mean relative humidity is 90\,\%. The substellar point is
  at 0$^{\circ}$ latitude and 180$^{\circ}$ longitude.}
\label{fig:rhModelTest}
\end{center}
\end{figure}

\begin{figure}
\begin{center}
\includegraphics[angle=0, width=14cm]{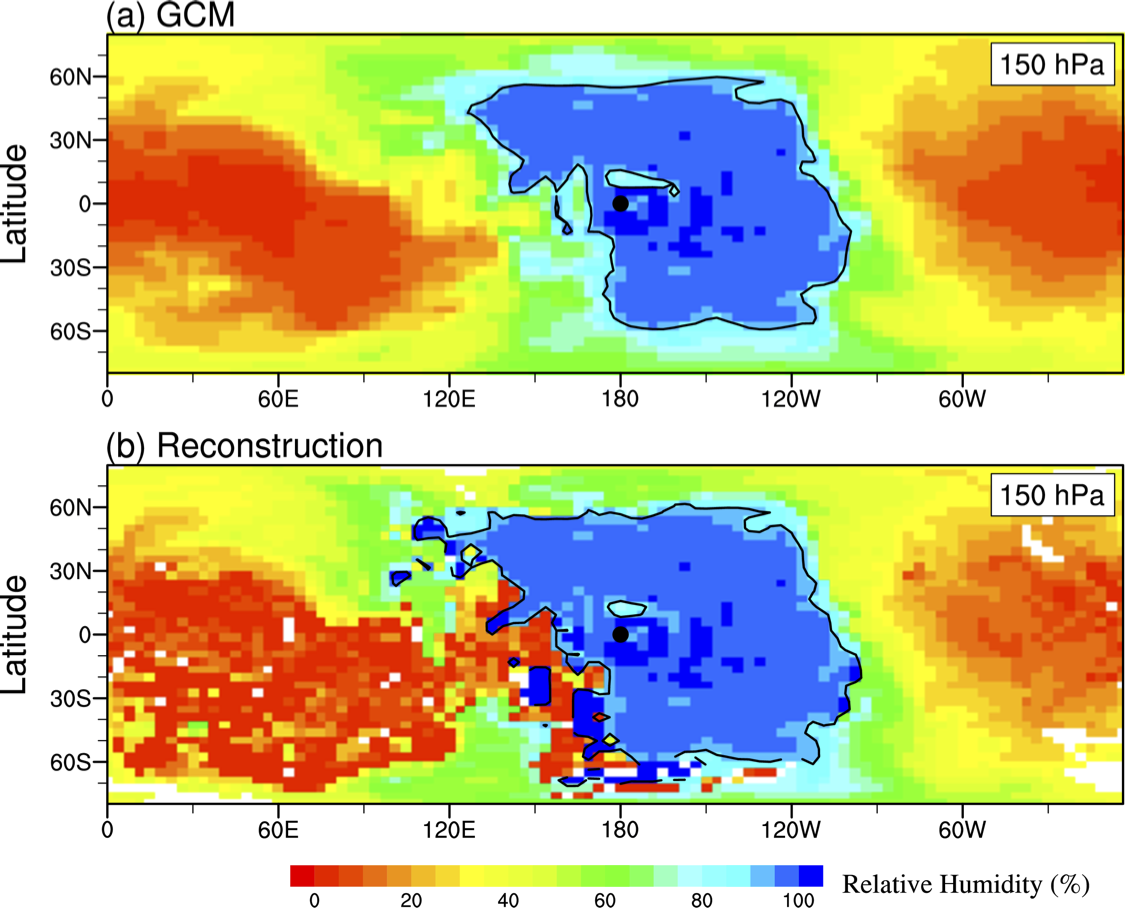}
\caption{Here the high-altitude (150~hPa) relative humidity from GCM 
  (a) is compared to the reconstructed relative humidity using the
  last saturation model (b) in a model snapshot. The white spots in
  the last saturation model represent areas that no air parcel we
  traced ended up at this particular time snapshot. The model is run
  in tidally locked aqua-planet configuration and forced with an M-star spectral
  energy distribution. The black dot is the substellar point, and the 
  black line is the contour of a relative humidity of 90\,\%.}
\label{fig:LastSatExample}
\end{center}
\end{figure}

This appendix briefly describes the last saturation model we built for
relative humidity in the CAM3 GCM based on the method of
\citet{pierrehumbert2007relative}. The model works by tracking an air
parcel as it moves around the planet and assuming that its specific
humidity is conserved after it reaches saturation for the last
time (Fig.~\ref{fig:LastSatModel}). This model does not include 
processes such as numerical
diffusion that can occur in the GCM. In a tidally locked planet, last
saturation generally occurs when convective ascent at the substellar
point ceases. The air parcel then flows away from the substellar point
and descends as it cools radiatively and heats adiabatically. We can
see these processes occuring in Fig.~\ref{fig:rhModelTest}, which
shows the trajectory of an example parcel and the relative humidity we
can infer for it using the last saturation method. The last saturation
model is able to reproduce the broad pattern of high-altitude relative
humidity as simulated by CAM3 (Fig.~\ref{fig:LastSatExample}). In
particular, the model reproduces the low relative humidity on the night side,
with approximately the correct magnitude. As would be expected for a
model without numerical diffusion, the relative humidity field from
the last saturation model is somewhat more noisy than that from CAM3.

\clearpage


\bibliography{intercomparison_20191224.bbl}

\end{document}